\documentclass[prd,byrevtex
,preprint
,showkeys
,showpacs
,amsfonts,amsmath,amssymb,
superscriptaddress
]{revtex4}
\usepackage[pdftex]{graphicx}
\usepackage{bm}
\usepackage{color}
\usepackage{amsthm}
\theoremstyle{definition}

\begin{document}
\title{Violation of Bell-CHSH inequalities through optimal local filters in the vacuum }
\author{Akira Matsumura}
\email{matsumura.akira@phys.kyushu-u.ac.jp}
\affiliation{Department of Physics, Kyushu University, Fukuoka, 819-0395, Japan }
\author{Yasusada Nambu }
\email{nambu@gravity.phys.nagoya-u.ac.jp}
\affiliation{Department of Physics, Graduate School of Science, Nagoya University, Chikusa, Nagoya 464-8602, Japan}
\date{September 28, 2020} %
%
\begin{abstract}
We investigate quantum correlations appearing for two qubit detectors which are initially uncorrelated and locally coupled to a massless scalar field in a vacuum state. Under the perturbation up to the second order in the coupling, the state of the detectors can be entangled through the interaction with the scalar field but satisfies the Bell-CHSH inequality. The violation of the Bell-CHSH inequality for such an entangled state is revealed by local filtering operations. In this paper, we construct the optimal filtering operations for the qubit detectors and derive the success probability of the filtering. The success probability characterizes the reliability of revealing the violation of the Bell-CHSH inequality by the filtering operations. Through these analyses, we demonstrate a trade-off relation between the success probability and the size of parameter region showing the violation of the Bell-CHSH inequality.
\end{abstract}
\maketitle

\tableofcontents
\section{Introduction}

In quantum information theory, quantum entanglement is known as a crucial
property which describes a nonlocal correlation \cite{Nielsen2007}. 
As well as quantum entanglement, the Bell-CHSH inequality also characterizes a nonlocal correlation in quantum mechanics \cite{Bell1964}. The Bell-CHSH inequality~\cite{Clauser1969} is
satisfied for local hidden variable theories, and a quantum
state violating this inequality cannot be described in that framework. The two aspects of quantum correlations,
quantum entanglement and the violation of the Bell-CHSH inequality, are not equivalent, and it was shown that a given quantum state violating the Bell-CHSH inequality has quantum entanglement~\cite{Werner1989}. 

In quantum field theory, it is known that quantum correlations appear in various states of a quantum field. Reeh and Schlieder showed that arbitrary
states of a quantum field can be approximated by acting some local
operators on the Minkowski vacuum state~\cite{Reeh1961}, and such a property
implies that the vacuum state is spatially entangled. Furthermore, it was shown that the vacuum state violates the Bell-CHSH inequality by examining the correlation of bounded observables between two spatial separated regions~\cite{Summers1985}. The vacuum of a quantum field
displays quantum entanglement and the violation of the Bell-CHSH
inequality, and these correlations characterize nonlocal features of quantum field.

In connection with the quantum correlations of the vacuum of a quantum field, the
detection of such correlations by local observers has been investigated. The so-called detector model is adopted to characterize how quantum resources in the vacuum is available, and provides a suitable experimental setting to detect the quantum correlation of the vacuum. The detector model is often described by spatially localized harmonic oscillators \cite{Lin2006,Brown2013a,Brown2013b} or qubit
systems~\cite{Reznik2005,Retzker2005, Silman2007,Leon2008, Salton2015,
  Pozas-Kerstjens2015,Sachs2017,Pozas-Kerstjens2017}. Reznik
\textit{et. al}~\cite{Reznik2005} considered the detection of quantum correlation by 
two qubit detectors. These detectors are locally coupled to a
massless scalar field and do not interact directly with each
other. It was shown that quantum entanglement can be extracted
and the violation of the Bell-CHSH inequality was found by applying a local filtering~\cite{Gisin1996}. The local filtering operation is a kind of measurement processes acting on each qubit by local observers Alice and Bob. The operation is constructed by
post-selected (probabilistic) local operations and classical
communication (LOCC). The proper choice of the operation reveals the violation of 
the Bell-CHSH inequality of the state of the detectors. This
method is also applied to the cosmological situation to reveal the
quantum nonlocality in the early universe~\cite{Nambu2011}. In quantum information theory, the optimal construction of local
filters which give the maximal violation of the Bell-CHSH inequality
was provided in the previous works~\cite{Verstraete2001,Verstraete2002}. 

In this paper, we investigate the quantum correlations in a vacuum through the model of two qubit detectors. 
The initial state of the detectors is usually assumed to be an uncorrelated ground state, however we also consider the excited state of detectors. By such a
generalization of the initial state of the detectors, we clarify what is playing a
crucial role to reveal the quantum entanglement and the violation of Bell-CHSH
inequality in the vacuum. As an entanglement measure, we compute the negativity of
the qubit detectors, which completely characterizes the entanglement
for two qubits system~\cite{Vidal2002a}. Also we construct the optimal
filtering operation for the two detectors by the method
given in Ref. \cite{Verstraete2002} to reveal the violation of the
Bell-CHSH inequality. We show that the constructed local filtering corresponds to that given previously in
Ref. \cite{Reznik2005} and derive the explicit formula of the success
probability of the filtering operation.  
From the formulas of quantum correlations and the success probability, it is shown that  the quantum correlations between the
detectors decreases and the success probability of the optimal
filtering increases as the transition probability
of the spontaneous emission grows. This behavior means that there is a trade-off relation between the
size of the parameter region indicating the quantum correlation
and the success probability.

This paper is organized as follows. In Sec. II, we introduce the system
composed of two qubit detectors and a massless scalar field. Up to the
second order in the coupling, we solve the dynamics assuming that the initial state of the total system is a product state of the detectors and the field in the Minkowski vacuum. We obtain the reduced density matrix
of the detectors represented by an X state. In Sec. III, we calculate the
negativity and the expectation value of the Bell operator for an X
state.  In Sec. IV, we explicitly construct the optimal filtering for
an X state and derive the success probability of the filtering.  In Sec
V, we discuss the quantum entanglement and the violation of the Bell-CHSH inequality
of detectors' system and show the quantum correlation is determined by
the coherence and the spontaneous emission of scalar
particles. In Sec. VI, we discuss the effect of local emissions to the violation of the Bell-CHSH inequality and find the trade-off relation between the parameter region revealing the violation of the Bell-CHSH inequality and the success probability of the optimal filter. Sec. VI is devoted to summary and conclusion.

\section{Perturbative dynamics of two detectors coupled to scalar field}

\begin{figure}[htbp]
\centering
    \includegraphics[clip,width=0.6\linewidth]{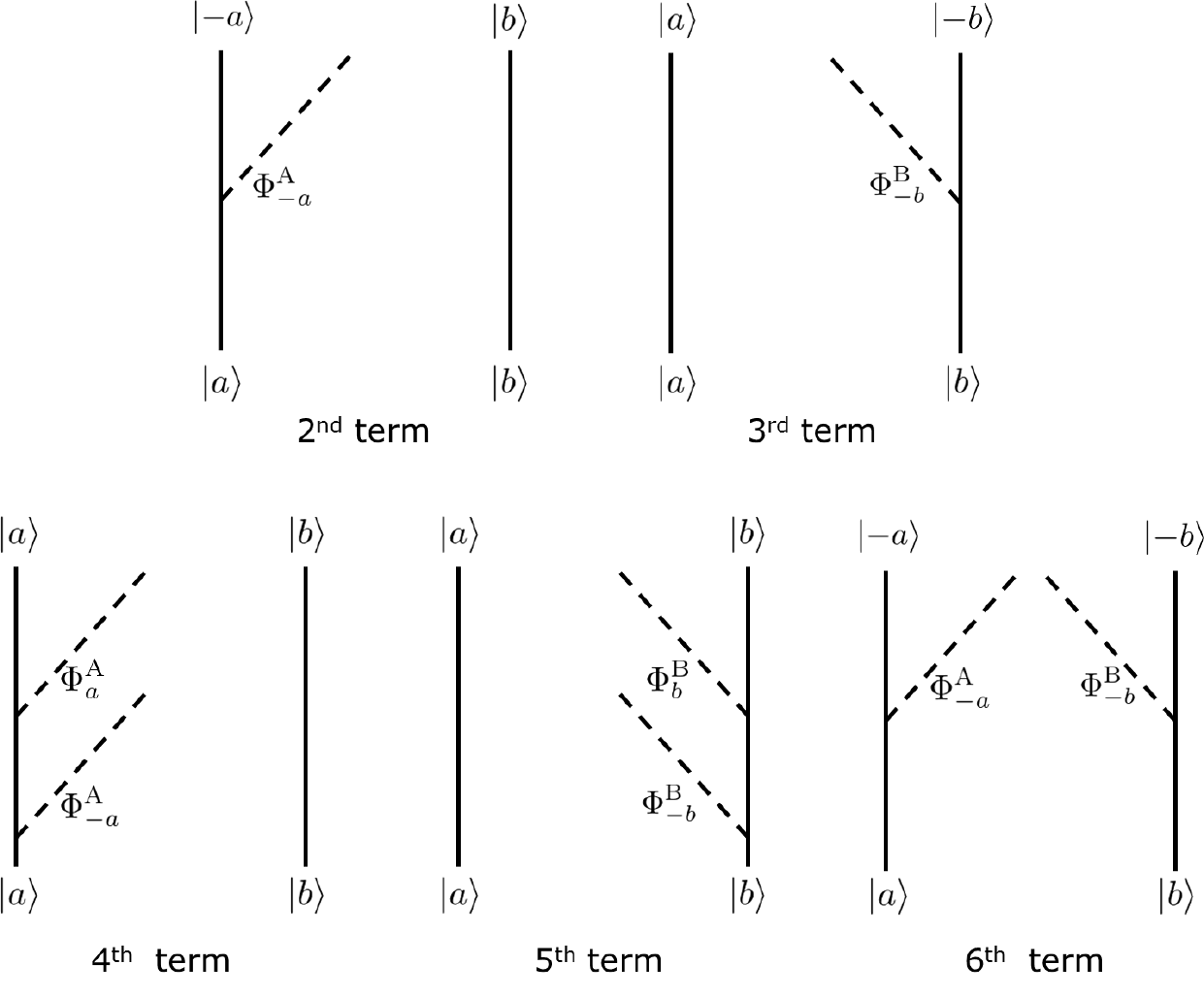}
    \caption{The diagrammatic representation of each term appeared in Eq. \eqref{eq5}.}
    \label{fig1}
\end{figure}

The vacuum state of a quantum field has nonlocal quantum correlations \cite{Reeh1961, Summers1985}. To investigate
the extraction of the quantum correlations by local observers, we consider a free theory of a massless scalar field and introduce qubit detectors locally coupled to the scalar field. The free Hamiltonian of the total system is $H_\text{free}=H_\text{A}+H_\text{B}+H_\phi$ with
\begin{equation}
H_\text{A}=\frac{\Omega}{2}\,\sigma^{z}_\text{A}, \quad
H_\text{B}=\frac{\Omega}{2}\,\sigma^{z}_\text{B}, \quad  H_{\phi}=\frac{1}{2} \int d^{3}x \bigl(\pi^{2}({\bm{x}})+(\nabla\phi({\bm{x}}))^{2} \bigr), 
\end{equation}
where $\sigma^{z}_\text{A,B}$ is the Pauli matrix, $\Omega$ is
the energy gap of the qubits,  $H_{\phi}$ is the free Hamiltonian of
the massless scalar field $\phi$ and $\pi:=\partial_{t}\phi$ is the
conjugate momentum of the scalar field. The interaction
Hamiltonian is
\begin{equation}
V(t)=g(t)\Bigl[\sigma_\text{A}^{x}\,\phi(\bm{x}_\text{A})+\sigma_\text{B}^{x}\,\phi(\bm{x}_\text{B})
\Bigr], \label{eq2}
\end{equation}
where $\bm{x}_\text{A}$ and $\bm{x}_\text{B}$ denote each spatial position of
the two detectors, that is, the two detectors are at rest at each
position and locally interact with the scalar field. We assume that
the switching function $g(t)$ has the Gaussian form
\begin{equation}
g(t)=g_{0} \exp \Bigl[ - \frac{(t-t_{0})^{2}}{2\sigma^{2}}\Bigr],
\end{equation}
where $g_{0}$ is a coupling constant and $\sigma$ is a time interval
while the interaction turns on. Roughly speaking, the detectors
interact with the scalar field for $|t-t_{0}| \leq \sigma$.
We assume that the initial state of the total system is a product state
\begin{equation}
|\Psi_{\rm{in}}\rangle=|a, b\rangle |0_{\phi}\rangle,
\end{equation}
where $a,b=\pm1$ denote eigenvalues of $\sigma^z_\text{A,B}$ and
$|0_{\phi} \rangle$ is the vacuum state of the scalar field. We
  also use the notation
  $|\!\uparrow\rangle=|\!+\!1\rangle, |\!\downarrow\rangle=|\!-\!1\rangle$ to
  represent the state of qubits. In the interaction picture, the out-state
  up to the second order in the coupling is given by
\begin{align}
|\tilde{\Psi}_{\rm{out}}\rangle
&\approx \Bigl[\mathbb{I}-i\int^{\infty}_{-\infty} dt_{1} \tilde{V}(t_{1})-\frac{1}{2}\int^{\infty}_{-\infty} dt_{1}\int^{\infty}_{-\infty} dt_{2}\, {\rm{T}}[\tilde{V}(t_{1})\tilde{V}(t_{2})] \Bigr] |\tilde{\Psi}_{\rm{in}} \rangle \nonumber \\
&=|a, b\rangle |0^{\phi} \rangle  -i|\!-\!a, b\rangle\,
  \Phi^\text{A}_{-a}|0_{\phi} \rangle-i |a, -b\rangle\,
  \Phi^\text{B}_{-b}|0^{\phi}\rangle \nonumber \\
& -\frac{1}{2}|a, b\rangle\,
  {\rm{T}}[\Phi^\text{A}_{a}\,\Phi^\text{A}_{-a}] |0_{\phi} \rangle
  -\frac{1}{2}|a, b\rangle\,
  {\rm{T}}[\Phi^\text{B}_{b}\,\Phi^\text{B}_{-b}] |0_{\phi} \rangle
  -|\!-\!a, -b\rangle\,  {\rm{T}}[\Phi^\text{A}_{-a}\Phi^\text{B}_{-b}] |0^{\phi}\rangle,  \label{eq5}
\end{align}
where $\tilde{V}$ is the interaction Hamiltonian in the interaction
picture, $\mathrm{T}$ denotes the time ordering, and the operators
$\Phi^\text{A}_{a}$ and $\Phi^\text{B}_{b}$ acting on the state of the scalar
field are defined by
\begin{equation}
\Phi^\text{A}_{a}=\int^{\infty}_{-\infty}dt\, g(t)\, e^{i\Omega at}\,
\phi({\bm{x}}_\text{A},t),\quad
\Phi^\text{B}_{b}=\int^{\infty}_{-\infty}dt\, g(t)\, e^{i\Omega bt}\, \phi({\bm{x}}_\text{B},t).
\end{equation}
Each term in Eq. \eqref{eq5} can be interpreted using the
diagrammatic representation shown in Fig. \ref{fig1}. For example, the second term in Eq.
\eqref{eq5} denotes that the  detector A interacts once with the
scalar field, and the qubit A is flipped.
\noindent
The reduced density matrix of the
two detectors after the interaction is 
\begin{align}
\rho_\text{AB}=
\begin{bmatrix}
     E_{0}(a,b)  & 0 & 0 & X(a,b)  \\
     0 & E_\text{A}(a) & E_\text{AB}(a,b)& 0 \\
     0 & E_\text{AB}^{*}(a,b) & E_\text{B}(b) & 0 \\
     X^{*}(a,b) &0 &0 & X_{4}(a,b)  
   \end{bmatrix}, \label{eq7}
\end{align}
where we assumed the basis 
$\{|1 \rangle,|2 \rangle,|3 \rangle,|4 \rangle \} = \{|a, b
\rangle,|\!\!-\!\!a, b\rangle,|a, -b\rangle,|\!-\!a, -b \rangle \}$.
The density matrix with only non-diagonal components $\rho_{23}:=\langle 2|\rho_\text{AB}|3 \rangle$ and $\rho_{14}:=\langle 1|\rho_\text{AB}|4 \rangle$ is called an X state, which has only the quantum coherences of the superpositions $\{|1\rangle, |4\rangle \}$
or $\{|2\rangle, |3\rangle \}$ and this property makes the
analysis of the quantum correlations easier. Concretely, the
non-diagonal components of the density matrix \eqref{eq7} are
\begin{align}
E_\text{AB}(a,b)&=\int d^{3}k \langle -a,b, \bm{k}_\phi | \tilde{\Psi}_\text{out} \rangle \langle \tilde{\Psi}_\text{out}| a,-b, \bm{k}_\phi \rangle =\langle 0_{\phi}| \Phi^\text{B}_{b}\,\Phi^\text{A}_{-a}|0_{\phi}\rangle, \label{eq:EAB} \\
X(a,b)&=\langle -a,-b, 0_{\phi}| \tilde{\Psi}_\text{out} \rangle=-\langle 0_\phi | {\rm{T}}[\Phi^\text{A}_{-a}\Phi^\text{B}_{-b}]^{\dagger} |0_{\phi}\rangle, \label{eq:X}
\end{align}
where $|\bm{k}_\phi \rangle$ is the one-particle state for the scalar field. The diagonal components are given as 
\begin{align}
E_{0}(a,b)&=1-E_\text{A}(a)-E_\text{B}(b)-X_{4}(a,b), \\
E_\text{A}(a)&=E_\text{AB}(a,a)|_{r=0}, \\
E_\text{B}(b)&=E_\text{AB}(b,b)|_{r=0}, \\
X_{4}(a,b)&=E_\text{A}(a) E_\text{B}(b)+|E_\text{AB}(a,b)|^{2}+|X(a,b)|^{2} \label{eq13}, 
\end{align}
where $r=|\bm{x}_\text{A}-\bm{x}_\text{B}|$ and the formula of
$X_{4}(a,b)$ is derived by the Wick theorem. Note that the
non-diagonal components $E_\text{AB}(a,b)$ and $X(a,b)$ depends
on the Wightman function for the massless scalar field
\begin{equation}
\langle 0_{\phi}|\phi(\bm{x}_\text{A},t)\phi(\bm{x}_\text{B},t')|0_{\phi} \rangle=-\frac{1}{4\pi^{2}}\frac{1}{(t-t'-i\epsilon)^{2}-r^{2}}, \label{eq13}
\end{equation}
where $\epsilon$ is the UV cutoff parameter. Hence the detectors with
  an initial product state can be entangled by the local interaction
with the scalar field in Eq. \eqref{eq2} through the two-point function. We can explicitly
compute $E_\text{A}(a), E_\text{B}(b), E_\text{AB}(a,b)$ and
$X(a,b)$ as
\begin{align}
E_\text{A}(a)&=\frac{g_{0}^{2}}{4\pi}  \bigl( e^{-(\Omega \sigma)^{2}}+2a\Omega \sigma\, {\mathrm{Erfc}}[-a\Omega \sigma] \bigr), \label{eq14}  \\
E_\text{B}(b)&=\frac{g_{0}^{2}}{4\pi}  \bigl( e^{-(\Omega \sigma)^{2}}+2b\Omega \sigma\, {\mathrm{Erfc}}[-b\Omega \sigma] \bigr), \label{eq15} \\
E_\text{AB}(a,b)&=\frac{g_{0}^{2}\sigma}{4\pi i r}e^{i\Omega(a-b)t_{0}-(\Omega\sigma)^{2}} \Bigl(\exp \Bigl[\bigl(-\frac{\Omega \sigma}{2}(a+b)-i\frac{r}{2\sigma}\bigr)^{2} \Bigr] {\rm{Erfc}} \Bigl[-\frac{\Omega \sigma}{2}(a+b)-i\frac{r}{2\sigma} \Bigr] \nonumber \\
&-\exp \Bigl[\bigl(-\frac{\Omega \sigma}{2}(a+b)+i\frac{r}{2\sigma}\bigr)^{2} \Bigr] {\rm{Erfc}} \Bigl[-\frac{\Omega \sigma}{2}(a+b)+i\frac{r}{2\sigma} \Bigr] \Bigr) \label{eq16} \\
X(a,b)&=\frac{g_{0}^{2} \sigma}{4\pi ir}e^{i\Omega (a+b)t_{0}-(\Omega \sigma)^{2}}   
\Bigl(\exp \Bigl[\Bigl(\frac{\Omega \sigma}{2}(a-b)-i\frac{r}{2\sigma} \Bigr)^{2}\Bigr] {\rm{Erfc}} \Bigl[\frac{\Omega \sigma}{2}(a-b)-i\frac{r}{2\sigma}\Bigr] \nonumber \\
&+\exp \Bigl[\Bigl(-\frac{\Omega \sigma}{2}(a-b)-i\frac{r}{2\sigma}\Bigr)^{2}\Bigr] {\rm{Erfc}} \Bigl[-\frac{\Omega \sigma}{2}(a-b)-i\frac{r}{2\sigma} \Bigr] \Bigr), \label{eq17}
\end{align}
where ${\rm{Erfc}}[z]$ is the complementary error function defined by 
\begin{equation}
{\rm{Erfc}}[z]=\int^{\infty}_{0}dt\, e^{-(t+z)^{2}}.
\end{equation}
The detailed derivation \eqref{eq16} and \eqref{eq17} is
presented in the Appendix A. From the explicit formulas for the
density matrix, the quantum correlation of the scalar field
detected via the
two detectors can be computed. 

\section{Negativity and  Bell-CHSH inequality for X state}

\begin{figure}[htbp]
\centering
    \includegraphics[clip,width=0.45\linewidth]{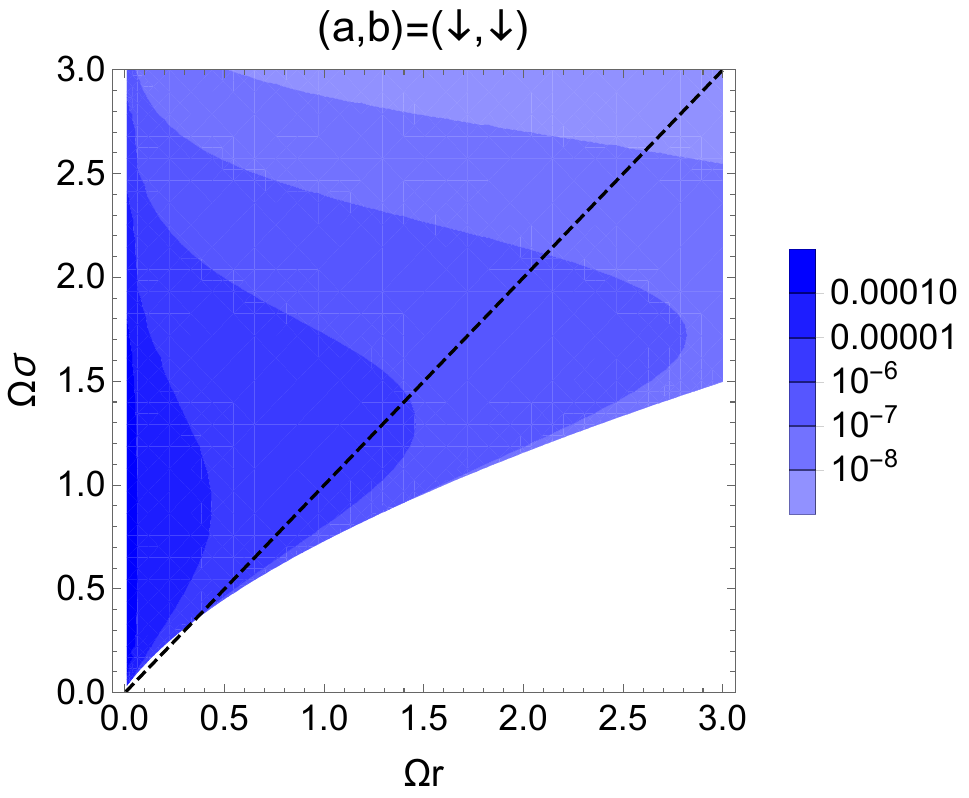}
    \caption{The contour plot of the negativity in the parameter space
        $(\Omega r, \Omega\sigma)$  with the initial detectors'
      state $|\downarrow_{A} \downarrow_{B} \rangle$.  The dashed line
      represents the null curve $r=\sigma$.}
    \label{fig2}
\end{figure}

As the state of
the detectors depends on the two-point function for the scalar field,
we expect that the initial product state of the detectors
becomes correlated after the interaction. To evaluate the
quantum correlation between the two detectors, we consider the
negativity and the Bell-CHSH inequality. The negativity is defined by
the eigenvalues,
$\lambda_i$, of a partial transposed density matrix
$\rho_\text{AB}^{\rm{T_{A}}}$ as 
\begin{equation}
\mathcal{N}=\sum_{\lambda_{i}<0} |\lambda_{i}|.
\end{equation}
If the negativity does not
vanish, then the state is entangled. Especially, the converse the
statement is true when the Hilbert space
$\mathcal{H}_\text{A} \otimes \mathcal{H}_\text{B}$ is
$\mathbb{C}^{2} \otimes \mathbb{C}^{2}$ or
$\mathbb{C}^{2} \otimes \mathbb{C}^{3}$~\cite{Horodecki1996}. Thus, the negativity completely characterizes whether the state of the detectors is entangled or not. For an X state
\begin{align}
\rho_\text{AB}=
\begin{bmatrix}
     \rho_{11}  & 0 & 0 & \rho_{14}  \\
     0 & \rho_{22} & \rho_{23} & 0 \\
     0 & \rho_{23}^{*} & \rho_{33} & 0 \\
     \rho_{14}^{*} &0 &0 & \rho_{44}  
   \end{bmatrix}, \label{}
\end{align}
the negativity is explicitly obtained as
\begin{align}
\mathcal{N}&=\max  \bigl[ \mathcal{N}_{1}, 0 \bigr]+\max \bigl[\mathcal{N}_{2},0 \bigr], \\
\mathcal{N}_{1}&=\frac{1}{2} \Bigl( \sqrt{(\rho_{11}-\rho_{44})^{2}+4|\rho_{23}|^{2}}-(\rho_{11}+\rho_{44})\Bigr), \\
\mathcal{N}_{2}&=\frac{1}{2} \Bigl( \sqrt{(\rho_{22}-\rho_{33})^{2}+4|\rho_{14}|^{2}}-(\rho_{22}+\rho_{33})\Bigr).
\end{align}
The conditions $\mathcal{N}_{1}>0$ or $\mathcal{N}_{2}>0$ are
rewritten in the simple form as
\begin{equation}
\sqrt{\rho_{11}\rho_{44}}< |\rho_{23}|\quad{\text{or}}\quad\sqrt{\rho_{22}\rho_{33}}< |\rho_{14}|. \label{eq25}
\end{equation}
For the detectors' density matrix \eqref{eq7}, the first inequality in \eqref{eq25} is not satisfied because we find 
\begin{equation}
\sqrt{\rho_{11}\rho_{44}}-|\rho_{23}| \sim \sqrt{E_\text{A}E_\text{B}+|E_\text{AB}|^{2}+|X|^{2}}-|E_\text{AB}| \geq 0 \label{}
\end{equation}
to the leading order in the coupling 
$g_0$, where 
$\rho_{11}=E_0$, 
$\rho_{44}=X_4$ and 
$\rho_{23}=E_\text{AB}$. 
Hence we can obtain the condition of nonzero negativity for the detectors' density matrix as 
\begin{equation}
\sqrt{E_\text{A}E_\text{B}} < |X|. \label{eq:ent-cond}
\end{equation}
For the detailed understanding of the quantum correlation,
it is important to evaluate the Bell-CHSH
inequality~\cite{Clauser1969} given by the correlation function for
the qubit A and B. To compute the Bell-CHSH inequality, we introduce
the Bell operator
\begin{equation}
\mathcal{B}_\text{AB}=\frac{1}{2} \bm{n} \cdot \bm{\sigma}_\text{A} \otimes (\bm{m}+\bm{m'}) \cdot \bm{\sigma}_\text{B}+\frac{1}{2} \bm{n'} \cdot \bm{\sigma}_\text{A} \otimes (\bm{m}-\bm{m'}) \cdot \bm{\sigma}_\text{B},
\end{equation}
where 
$\bm{n}, \bm{n'}, \bm{m}$ and $\bm{m'}$ are unit vectors, 
$\bm{\sigma}_\text{A}$ and 
$\bm{\sigma}_\text{B}$ are the Pauli matrices. We consider
the maximum expectation value $\beta$ of the Bell-CHSH
operator
\begin{equation}
\beta(\rho_\text{AB})=\max_{\bm{n}, \bm{n'}, \bm{m},\bm{m'}} 
{\rm{Tr}}[\mathcal{B}_\text{AB}\,\rho_\text{AB}].
\end{equation}
For separable states, $\beta(\rho_\text{AB})$ satisfies
 the following Bell-CHSH inequality
\begin{equation}
\beta(\rho_\text{AB}) \leq 1. \label{eq26}
\end{equation}
The inequality \eqref{eq26} holds for the local hidden variable theory, which includes any separable states. For any physical states,
$\beta(\rho_\text{AB})$ has the upper bound called the Tsirelson
bound~\cite{Tsirelson1980}
\begin{equation}
\beta(\rho_\text{AB}) \leq \sqrt{2}.
\end{equation}
For an X state, the maximum value $\beta(\rho_\text{AB})$ can be calculated
explicitly as
\begin{align}
\beta(\rho_\text{AB})&=\max [\beta_{1},\beta_{2}], \label{eq:beta} \\
\beta_{1}&=\sqrt{(\rho_{11}+\rho_{44}-\rho_{22}-\rho_{33})^{2}+4(|\rho_{14}|+|\rho_{23}|)^{2}}, \\
\beta_{2}&=2\sqrt{(|\rho_{14}|+|\rho_{23}|)^{2}+(|\rho_{14}|-|\rho_{23}|)^{2}},
\end{align}
where we used the Horodecki theorem \cite{Horodecki1995} which
provides the method to obtain the explicit form of $\beta$ from the
singular value of the matrix
$R^{ij}={\mathrm{Tr}}[\sigma^{i}_\text{A}\sigma^{j}_\text{B}\,\rho_\text{AB}]$. Note
that the Bell-CHSH inequality is satisfied for the state of the two detectors system given by (15)-(18) within our perturbative treatment. Since the order of the coupling 
$g_{0}$ for the non-diagonal
components 
$E_\text{AB}$ and 
$X$ is 
$O(g^{2}_{0})$,
$\beta_{1}$ and 
$\beta_{2}$ for a small 
$g_{0}$ are evaluated as
\begin{equation}
\beta_{1}=1-2(E_\text{A}+E_\text{B})+O(g_0^4),\quad\beta_{2}= O(g^{2}_{0}),
\end{equation}
where $E_\text{A}$ and $E_\text{B}$ are $O(g_{0}^{2})$. The
maximum expectation value of the Bell operator $\beta$ is smaller than
unity and the Bell-CHSH inequality is always satisfied. On the other
hand, it is possible for the detectors to have a nonzero negativity
because the condition for the entangled state \eqref{eq:ent-cond} does not
depend on the strength of coupling (the both sides of the inequality
\eqref{eq:ent-cond} have the same order for the coupling). Fig.~\ref{fig2}
shows the contour plot of the negativity in $(\Omega r, \Omega\sigma)$
space for the detectors' initial state
$|\!\downarrow_{A} \downarrow_{B} \rangle$.  The dashed line denotes
the ``null'' curve $r=\sigma$ and we find that the negativity has
a nonzero value for a region 
$r>\sigma$.
\noindent
As we have seen above, the state of the detectors is entangled and
satisfies the Bell-CHSH inequality. Interestingly, it is known that
the violation of the Bell-CHSH inequality for
such a state can be revealed by a local filtering operation \cite{Gisin1996}.


\section{Local filtering operation for X states}

We introduce a local filtering operation for a two-qubit system. The local filtering operation is performed as follows: qubit locally interacts with a measurement apparatus and then an observer picks up one outcome from the measurement apparatus. When two local observers, Alice and Bob, perform the local filtering operation for each qubit, the filtered state 
$\rho'_\text{AB}$ is given by
\begin{equation}
  \rho'_\text{AB}= \frac{1}{p} M_\text{A} \otimes N_\text{B}\, \rho_\text{AB}\, M_\text{A}^{\dagger} \otimes N_\text{B}^{\dagger},
\label{eq:filtered}
\end{equation}
where 
$\rho_\text{AB}$ is an initial density operator of a two-qubit system, 
$M_{A}$ and $N_{B}$ are local operators ($2 \times 2$ matrices) for
each subsystem. The success probability, $p$, to attain the filtered state is   
\begin{equation}
p={\mathrm{Tr}}[M_\text{A}^{\dagger}M_\text{A}
\otimes N_\text{B}^{\dagger}N_\text{B} \,\rho_\text{AB}]. 
\label{eq:p}
\end{equation}
Those operators 
$M_\text{A}$ and 
$N_\text{B}$ have inverse and satisfy the
conditions
$M_\text{A}^{\dagger}M_\text{A} \leq \mathbb{I}_\text{A}$ and 
$N_\text{B}^{\dagger} N_\text{B} \leq \mathbb{I}_\text{B}$, where 
$\mathbb{I}_\text{A}$ and 
$\mathbb{I}_\text{B}$ are the identity operators on each Hilbert space of the qubit systems. It is known that the local filtering is adopted to reveal the violation of Bell-CHSH inequality for the bipartite qubit system \cite{Gisin1996,Verstraete2001,Verstraete2002}.

\subsection{Key theorems}
There are two important theorems to reveal the violation of Bell-CHSH
inequality by the local filtering operation
\cite{Verstraete2001,Verstraete2002}:

 \noindent
 \textbf{Theorem 1} \cite{Verstraete2001} By a local filtering
 operation,  a two-qubit state $\rho_\text{AB}$ can be uniquely
 transformed into a Bell diagonal state.

\noindent
\textbf{Theorem 2 }\cite{Verstraete2002}
  If the optimized $\beta(\rho'_\text{AB})$ over all local operations
  $M_\text{A}$ and $N_\text{B}$ is larger than unity, then the filtered state
  $\rho'_\text{AB}$ is a Bell diagonal state. The statement is
  represented by
\begin{equation}
\max_{M_\text{A},N_\text{B}} \beta( \rho'_\text{AB}) >1 \quad \Longrightarrow \quad \rho'_\text{AB}=\sum_{\mu=0}^{3} \lambda_{\mu} | \rm{Bell}^{\mu}_\text{AB} \rangle \langle  \rm{Bell}^{\mu}_\text{AB}|,
\end{equation} 
where
$|\rm{Bell}^{\mu}_{AB} \rangle:= \sigma^{\mu}_{A}
(|\uparrow_{A}\uparrow_{B} \rangle + |\downarrow_{A}\downarrow_{B}
\rangle)/\sqrt{2}$
and $\sigma^{\mu}=\{\mathbb{I}, \sigma^{x}, \sigma^{y}, \sigma^{z} \}$.

According to above theorems, we need the local operation which transforms
a given state into a Bell diagonal form to reveal the violation of Bell-CHSH inequality. In general, it is complicated to construct such a local operation, however we easily get it for an X state. We note that
a Bell diagonal state
$\sum_{\mu=0}^{3} \lambda_{\mu} | \rm{Bell}^{\mu}_{AB} \rangle \langle
\rm{Bell}^{\mu}_{AB}|$ has the form of an X state with its
  components given by
\begin{align}
\frac{1}{2}
\begin{bmatrix}
     \lambda_{0}+\lambda_{3}  & 0 & 0 & \lambda_{0}-\lambda_{3}  \\
     0 & \lambda_{1}+\lambda_{2} & \lambda_{1}-\lambda_{2}& 0 \\
     0 & \lambda_{1}-\lambda_{2} & \lambda_{1}+\lambda_{2} & 0 \\
     \lambda_{0}-\lambda_{3} &0 &0 & \lambda_{0}+\lambda_{3} 
   \end{bmatrix},
\quad \sum_{\mu=0}^3\lambda_\mu=1,\quad\lambda_\mu\ge 0.
 \label{eq:Belldiag}
\end{align}
This state corresponds to the X state with
\begin{equation}
\rho_{11}=\rho_{44}, \quad\rho_{22}=\rho_{33},
\quad\rho_{14}=\rho_{14}^{*},\quad\rho_{23}=\rho_{23}^{*}.
\label{eq:X-cond}
\end{equation}
All we have to do is to transform a given X state into the X state
satisfying these conditions \eqref{eq:X-cond} by an appropriate filtering
  operation.  We apply the local $z$ rotation
$\exp[-i\theta\,\sigma^{z}_\text{A}/2-i\phi\,\sigma^{z}_\text{B}/2]$
to a given X state. The diagonal components are invariant and the
non-diagonal components are transformed as
\begin{equation}
\rho_{14} \rightarrow e^{-i(\theta+\phi)} \rho_{14}, \quad \rho_{23} \rightarrow e^{-i(\theta-\phi)} \rho_{23}.
\end{equation} 
We can choose the parameters $\theta, \phi$ so that
$\rho_{14}, \rho_{23}$ are positive and satisfy
$\rho_{23}=\rho_{23}^{*}, \rho_{14}=\rho_{14}^{*}$. Without loss of
generality, we assume that the diagonal components satisfy
$\rho_{11} \geq \rho_{22} \geq \rho_{33} \geq \rho_{44}$.  From
  the theorem 1, we can uniquely transform the two qubit system to a
  Bell diagonal form by a local filtering operation. Hence it is
  sufficient to find one of the filtering operations converting a
  given X state to a Bell diagonal state. For this purpose, we consider the
local operation defined by
\begin{align}
M_\text{A}=
\begin{bmatrix}
     \eta_\text{A}  & 0 \\
     0 & 1 
   \end{bmatrix},\quad 
N_\text{B}=
\begin{bmatrix}
     \eta_\text{B}  & 0 \\
     0 & 1 
   \end{bmatrix}, 
\label{eq:lfilter}
\end{align}
where $0<\eta_\text{A}^{2} \leq1$ and $0<\eta_\text{B}^{2} \leq 1$.
This operation corresponds to the amplitude damping channel with a
post selection and was used in Ref. \cite{Reznik2005} to detect the
Bell-CHSH nonlocality.  Under the local operations \eqref{eq:lfilter},
the X state is transformed to
\begin{equation}
 \rho'{}_\text{AB}
=\frac{1}{p}
\begin{bmatrix}
  \eta_\text{A}^{2}\eta_\text{B}^{2}\,\rho_{11} & 0 & 0 &
  \eta_\text{A}\eta_\text{B}\,|\rho_{14}| \\
  0 & \eta^{2}_\text{A}\, \rho_{22}  & \eta_\text{A}\eta_\text{B}\,
  |\rho_{23}|& 0 \\ 
  0 &\eta_\text{A}\eta_\text{B}\,|\rho_{23}| & \eta^{2}_\text{B}\, \rho_{33}& 0 \\
  \eta_\text{A}\eta_\text{B}\,|\rho_{14}| &0 &0 & \rho_{44}
   \end{bmatrix},
\end{equation}
where
$p= \eta_\text{A}^{2}\eta_\text{B}^{2}\,\rho_{11}+\eta^{2}_\text{A}
\,\rho_{22}+\eta^{2}_\text{B}\, \rho_{33}+\rho_{44}$.
If the parameters $\eta_\text{A}$ and $\eta_\text{B}$ satisfy
$\eta_\text{A}^{2}\eta_\text{B}^{2}\,\rho_{11}=\rho_{44},~\eta^{2}_\text{A}
\rho_{22}=\eta^{2}_\text{B}\, \rho_{33}$, that is,
\begin{equation}
\eta_\text{A}^{2}= \Bigl(\frac{\rho_{44}\rho_{33}}{\rho_{11}\rho_{22}}
\Bigr)^{1/2},\quad \eta_\text{B}^{2}=
\Bigl(\frac{\rho_{44}\rho_{22}}{\rho_{11}\rho_{33}} \Bigr)^{1/2}, \label{eq:opt_filter}
\end{equation}
then the X state becomes the Bell diagonal state with the spectrum
$\{\lambda_{\mu}\}$ given by
\begin{align}
\lambda_{0}&=\frac{\sqrt{\rho_{11}\rho_{44}}+|\rho_{14}|}{2(\sqrt{\rho_{11}\rho_{44}}+\sqrt{\rho_{22}\rho_{33}})},\quad
\lambda_{1}=\frac{\sqrt{\rho_{22}\rho_{33}}+|\rho_{23}|}{2(\sqrt{\rho_{11}\rho_{44}}+\sqrt{\rho_{22}\rho_{33}})}, \nonumber \\
\lambda_{2}&=\frac{\sqrt{\rho_{22}\rho_{33}}-|\rho_{23}|}{2(\sqrt{\rho_{11}\rho_{44}}+\sqrt{\rho_{22}\rho_{33}})},\quad
\lambda_{3}=\frac{\sqrt{\rho_{11}\rho_{44}}-|\rho_{14}|}{2(\sqrt{\rho_{11}\rho_{44}}+\sqrt{\rho_{22}\rho_{33}})}. \label{eq:lambdas}
\end{align}
Eq.~\eqref{eq:opt_filter} provides the optimal filters for the detection of the violation of Bell-CHSH inequality with the success
probability $p$
\begin{equation}
p=2 \rho_{44} \Bigl[1+\Bigl(\frac{\rho_{22}\rho_{33}}{\rho_{11}\rho_{44}}\Bigr)^{1/2} \Bigr]. \label{eq46}
\end{equation}
This probability characterizes the reliability of detecting
the violation of Bell-CHSH inequality by the local filtering operation.

\subsection{Quantum correlation of Bell diagonal state and coherence
  of X state}

To get clear understanding of the quantum correlation for the
  X state, we investigate the detailed properties of the Bell
diagonal state and its relationship to the X state. The entanglement of
the Bell diagonal state is completely characterized by the
negativity. The conditions of non-zero negativity \eqref{eq25} for the
Bell diagonal state are equivalent to
\begin{equation}
(\lambda_{0}-1/2)(\lambda_{3}-1/2)<0\quad{\rm{or}}\quad(\lambda_{1}-1/2)(\lambda_{2}-1/2)<0, \label{eq42}
\end{equation}
where we used Eq. \eqref{eq:Belldiag}. Hence, whenever the largest
eigenvalue of $\lambda_{\mu}$ exceeds $1/2$ (the spectrum 
$\{ \lambda_{\mu} \}$ is biased towards any one of the
four Bell states), then the Bell diagonal state is entangled.

\begin{figure}[htbp]
\centering
    \includegraphics[clip,width=0.7\linewidth]{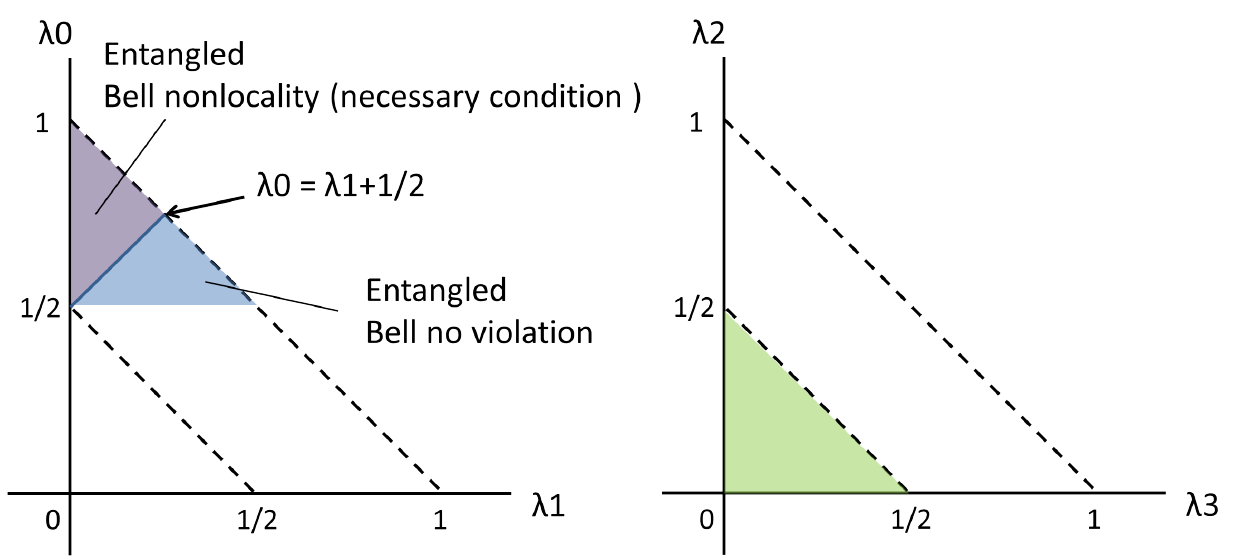}
    \caption{The typical region of $\{\lambda_{\mu}\}$ revealing the
      quantum entanglement and the violation of Bell-CHSH inequality.}
    \label{fig3}
\end{figure}

Let us focus on the violation of Bell-CHSH inequality for the Bell diagonal state. When the
maximum value $\beta$ is larger than $1$ (that is, $\beta_{1}>1$ or
$\beta_{2}>1$ in Eq. \eqref{eq:beta}), the eigenvalues 
 satisfy
\begin{equation}
(\lambda_{0}-\lambda_{2})^{2}+(\lambda_{1}-\lambda_{3})^{2}>1/2\quad{\text{or}}\quad
(\lambda_{0}-\lambda_{1})^{2}+(\lambda_{2}-\lambda_{3})^{2}>1/2, 
\end{equation}
where $\lambda_{0} \geq \lambda_{3}$ and
$\lambda_{1} \geq \lambda_{2}$ are imposed by Eq.~\eqref{eq:lambdas}. If we assume $\lambda_{0}>1/2$ then
$(\lambda_{1}-\lambda_{3})^{2} \leq1/4$ and
$(\lambda_{2}-\lambda_{3})^{2} \leq1/4$ because
  $\sum_\mu \lambda_\mu=1$.  We obtain the inequalities
\begin{equation}
\lambda_{0}-\lambda_{2}>\frac{1}{2}\quad \text{or}\quad\lambda_{0}-\lambda_{1}>\frac{1}{2} \label{eq44}
\end{equation}
as the necessary condition of the violation of the Bell-CHSH inequality. 
To summarize, the typical region of the spectra satisfying the
entanglement condition \eqref{eq42} and the violation of Bell-CHSH inequality
(necessary) conditions \eqref{eq44} are presented in
Fig.~\ref{fig3}. The Bell diagonal state has the
Bell-CHSH nonlocal correlation when  one of the spectra $\{\lambda_{\mu}\}$ 
approaches unity.

In Eq \eqref{eq:lambdas}, the spectra
$\{\lambda_{\mu}\}$ depend on the components of the X state and
their dominant terms are $|\rho_{23}|$ and $|\rho_{14}|$. To show the violation of the Bell-CHSH inequality, one of these terms, and associated coherence that it represents, need to dominate the other. 

\section{Violation of Bell-CHSH inequality by using optimal local filtering}

In this section, we examine the quantum entanglement and the
  violation of Bell-CHSH inequality for the model of two qubit detectors with the
  initial conditions
  $|\downarrow_\text{A} \downarrow_\text{B}\rangle, |\uparrow_\text{A}
  \uparrow_\text{B}\rangle$
  and $|\downarrow_\text{A} \uparrow_\text{B}\rangle$.  For the
detection of the violation of Bell-CHSH inequality, we apply the local filter to the
qubit detectors' state given in Sec.~IV.  

\subsection{The initial condition $|\downarrow_\text{A} \downarrow_\text{B} \rangle$}

\begin{figure}[htbp]
   \centering
   \includegraphics[clip, width=0.45\linewidth]{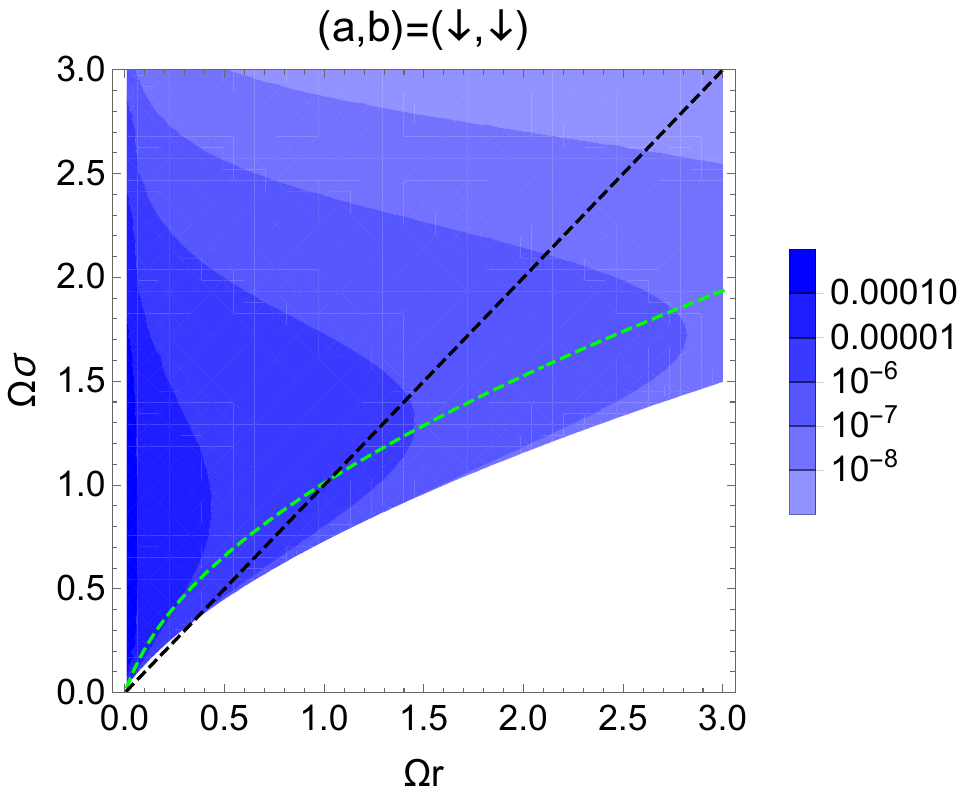}
 \caption{ The contour plot of the negativity with the
   initial condition $|\!\downarrow_\text{A}\downarrow_\text{B}\rangle$.
   The green dashed line denotes $\beta=1$, and the above this line
   $\beta>1$.}
    \label{fig4}
\end{figure}

We consider the initial condition of the detectors
$(a,b)=(-1,-1)$ corresponding to the state
$|\!\downarrow_\text{A} \downarrow_\text{B} \rangle$. From Eqs. \eqref{eq14},\eqref{eq15},\eqref{eq16} and \eqref{eq17}, we
derive
\begin{align}
E_\text{A}(-1)&=\frac{g_{0}^{2}}{4\pi}  \bigl( e^{-(\Omega \sigma)^{2}}-2\Omega \sigma\, {\mathrm{Erfc}}[\Omega \sigma] \bigr), \\
E_\text{AB}(-1,-1)&=\frac{g_{0}^{2}\sigma}{4\pi i r}e^{-(r/2\sigma)^{2}} \Bigl(e^{-i\Omega r}\, {\rm{Erfc}} \Bigl[\Omega \sigma-i\frac{r}{2\sigma} \Bigr] -e^{i\Omega r} \,{\rm{Erfc}} \Bigl[\Omega \sigma+i\frac{r}{2\sigma} \Bigr] \Bigr),  \\
X(-1,-1)&=\frac{g_{0}^{2} \sigma}{2\pi ir}e^{-2i\Omega t_{0}-(\Omega \sigma)^{2}-(r/2\sigma)^{2}}   
\,{\rm{Erfc}} \Bigl[-i\frac{r}{2\sigma}\Bigr],
\end{align}
and $E_\text{B}(-1)=E_\text{A}(-1)$. Fig.~\ref{fig4}
shows the contour plot of the negativity for the filtered X state
with the initial condition  $|\!\downarrow_\text{A} \downarrow_\text{B}
\rangle$. The coupling $g_{0}$ is fixed to $10^{-2}$.
\noindent
The green dashed line denotes $\beta=1$, and the region above
this line represents $\beta>1$ where the Bell-CHSH
inequality is violated. In addition, we observe the existence of
the region where the violation of Bell-CHSH inequality is not found even if the optimal
filtering is acted on each detector. 
   
\begin{figure}[htbp]
\centering
    \includegraphics[clip,width=0.5\linewidth]{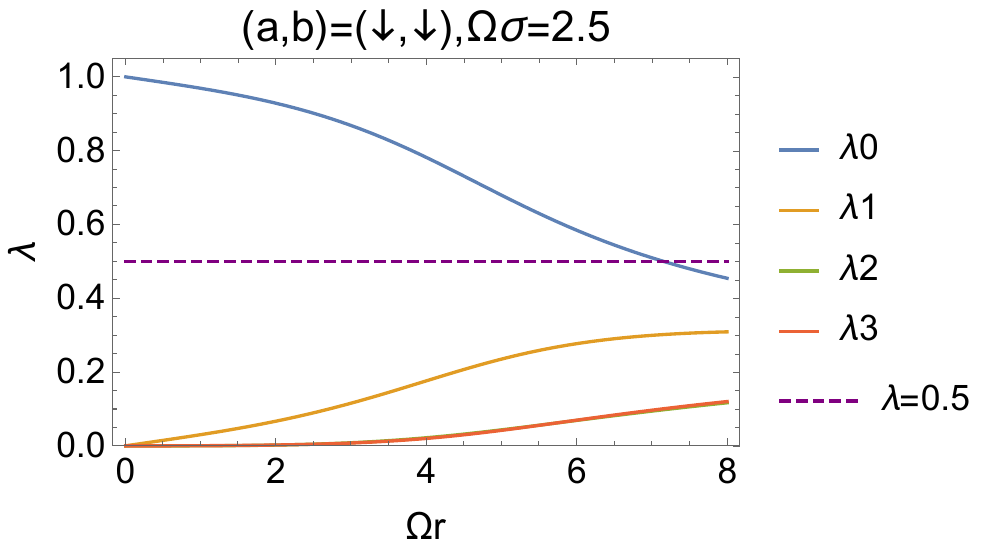}
    \caption{The behavior of the spectrum $\{\lambda_{\mu}\}$ of the Bell
      diagonal state with fixed $\Omega \sigma=2.5$ and
      $g_{0}=10^{-2}$. The initial condition of the detectors' state
      is $|\!\!\downarrow_\text{A}\downarrow_\text{B} \rangle$. $\lambda_{0}$ is
      larger than  the other eigenvalues, that is, the coherence
      $|X|$ is dominant.}
    \label{fig5}
\end{figure}

\begin{figure}[htbp]
\centering
    \includegraphics[clip,width=0.5\linewidth]{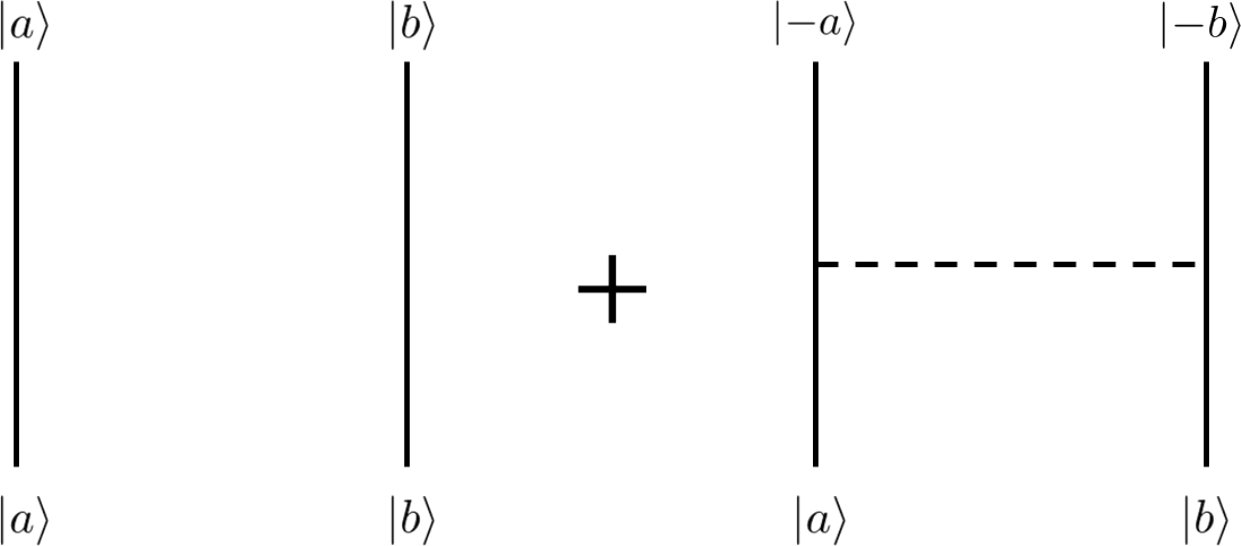}
    \caption{The diagrammatic picture of the coherence generation by the exchange of the scalar field.}
    \label{fig6}
\end{figure}

We analyze how the quantum correlation of the scalar field is
detected through the detectors. In Sec.~IV, we give the simple form \eqref{eq:lambdas} of
the spectrum $\{\lambda_{\mu}\}$ obtained from the components of the X
state. Fig.~\ref{fig5} shows the behavior of those
spectra with $\Omega \sigma=2.5$ and we observe that 
$\lambda_{0}$ is dominant compared to the others.
From Eq. \eqref{eq:lambdas} we note that
the eigenvalue $\lambda_{0}$ for a small coupling is evaluated as
\begin{equation}
\lambda_{0} \approx \frac{1}{2}+\frac{|X|-\sqrt{E_\text{A}E_\text{B}}}{2(\sqrt{E_\text{A}E_\text{B}+|E_\text{AB}|^{2}+X^{2}}+\sqrt{E_\text{A}E_\text{B}})}. \label{eq:lambda0}
\end{equation}
Hence the condition $\lambda_{0}>1/2$ is equivalent to
$|X|>\sqrt{E_\text{A}E_\text{B}}$, and Fig.\ref{fig5} means that the coherence $|X|$ of the superposition in the basis 
$\{ |\!\uparrow_\text{A} \uparrow_\text{B} \rangle, |\!\downarrow_\text{A}
\downarrow_\text{B} \rangle \}$
is larger than $\sqrt{E_\text{A}E_\text{B}}$. To understand why the coherence
$|X|$  dominates, we remind that the state of the detectors
depends on the two-point function of the scalar field. The
quantum superposition in the basis 
$\{ |\!\!\uparrow_\text{A}\uparrow_\text{B} \rangle,
|\!\!\downarrow_\text{A}\downarrow_\text{B} \rangle \}$
is realized by the exchange of the real or virtual scalar
field. Fig. \ref{fig6} corresponds to the diagrammatic picture to generate the coherence 
$|X|$ 
in the second order dynamics.

\subsection{The initial condition $|\!\uparrow_\text{A} \uparrow_\text{B} \rangle$}

We consider the detection of the quantum correlation
for the initial state $|\!\uparrow_\text{A}\uparrow_\text{B}\rangle$. The
components $E_\text{A}, E_\text{B}, E_\text{AB}$ and $X$ of the
reduced density matrix are
\begin{align}
E_\text{A}(+1)&=\frac{g_{0}^{2}}{4\pi}  \bigl( e^{-(\Omega \sigma)^{2}}+2\Omega \sigma \,{\rm{Erfc}}[-\Omega \sigma] \bigr), \\
E_\text{AB}(+1,+1)&=\frac{g_{0}^{2}\sigma}{4\pi i r}e^{-(r/2\sigma)^{2}} \Bigl(e^{i\Omega r}\, {\rm{Erfc}} \Bigl[-\Omega \sigma-i\frac{r}{2\sigma} \Bigr] -e^{-i\Omega r}\, {\rm{Erfc}} \Bigl[-\Omega \sigma+i\frac{r}{2\sigma} \Bigr] \Bigr),  \\
X(+1,+1)&=\frac{g_{0}^{2} \sigma}{2\pi ir}e^{2i\Omega t_{0}-(\Omega \sigma)^{2}-(r/2\sigma)^{2}}\, {\rm{Erfc}} \Bigl[-i\frac{r}{2\sigma}\Bigr],
\end{align}
and $E_\text{B}(+1)=E_\text{A}(+1)$. We find that
$|X(+1,+1)|=|X(-1,-1)|$, that is, those coherences with
the two different initial conditions are equivalent. Due to the facts that the vacuum is invariant under time translation and time reversal and that the switching function 
$g(t)$ has the symmetric property 
$g(t+t_0)=g(-t+t_0)$,  we can derive that the transition probability 
$|X(-1,-1)|^2$ of 
$|\!\downarrow_\text{A}\downarrow_\text{B} \rangle
\rightarrow |\!\uparrow_\text{A}\uparrow_\text{B} \rangle$ is the
same as 
$|X(+1,+1)|^2$ of
$|\!\uparrow_\text{A}\uparrow_\text{B} \rangle 
\rightarrow |\!\downarrow_\text{A}\downarrow_\text{B} \rangle$.
The detail calculation is presented in the Appendix B.

\begin{figure}[htbp]
   \centering
   \includegraphics[width=0.43\linewidth]{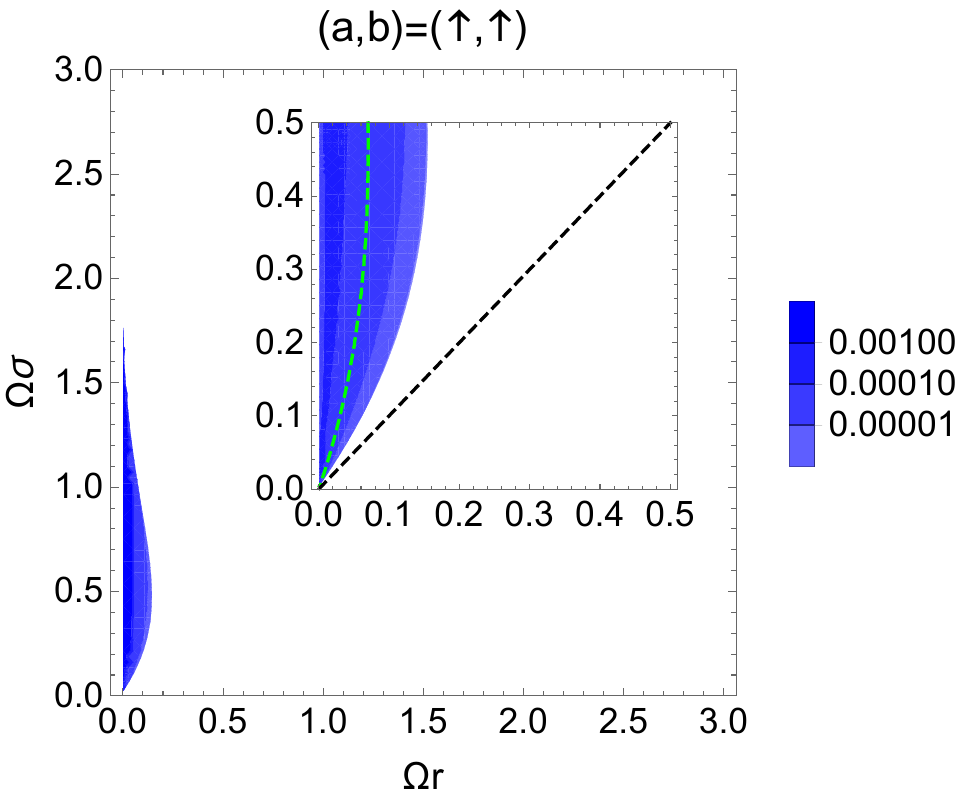}
   \includegraphics[width=0.5\linewidth]{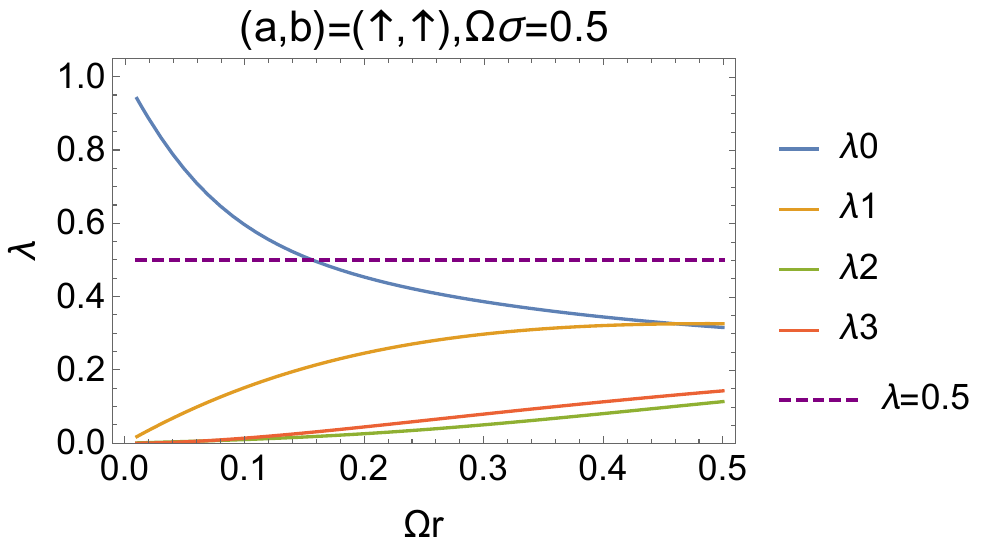}
 \caption{Left panel : The contour plot of the negativity with the
   initial condition $|\!\uparrow_{A}\uparrow_{B}\rangle$. The inset
   is the enlarged version of the contour plot. The green dotted line
   represents $\beta=1$.  $\mathcal{N}\neq 0$ region does not extend
   to the region $r>\sigma$.  Right panel: The behavior
   of the spectrum $\{\lambda_{\mu}\}$ of the Bell diagonal state with
   fixed $\Omega \sigma=0.5$ and $g_{0}=10^{-2}$.  The eigenvalue
   $\lambda_{0}$ is dominant due to the exchange of the scalar
   particle.}
    \label{fig7}
\end{figure}
The left panel of Fig. \ref{fig7} presents the contour plot of the
negativity with the initial state
$|\!\uparrow_\text{A}\uparrow_\text{B} \rangle$. The region with the
nonzero negativity is much smaller compared to the result
obtained with the initial state
$|\!\downarrow_\text{A}\downarrow_\text{B}\rangle$ (similar to the case for $(a,b)=(-1,-1)$, in the right panel of Fig. \ref{fig7}, we can
confirm that the quantum correlation of the X state is generated by the
exchange of the scalar particle because $\lambda_0$ dominates). To understand the different feature of quantum correlations from that for the case
$|\!\downarrow_\text{A}\downarrow_\text{B}\rangle$, we focus on the diagonal components $E_\text{A} (a)$ and
$E_\text{B}(b)$, which are the transition probabilities of 
$|a, b \rangle | \rightarrow |\!\!-\!a, b \rangle$
and
$|a, b \rangle  \rightarrow |a, -b \rangle $, respectively. Evaluating the difference $E_\text{A}(+1)-E_\text{A}(-1)$,
we obtain
\begin{equation}
E_\text{A}(+1)-E_\text{A}(-1)=\frac{g_{0}^{2}\Omega \sigma}{2\pi} \int^{\Omega \sigma}_{-\Omega \sigma} dt\, e^{-t^{2}} \geq 0.
\end{equation}
The inequality 
$E_\text{B}(+1) \geq E_\text{B}(-1)$ also holds. 
$E_\text{A}(+1)$ and $E_\text{B}(+1)$ correspond to the probabilities of spontaneous emissions, which are determined by the local dynamics and prevent the detection of nonlocal correlations (indeed, the equality  
$|X(-1,-1)|=|X(+1,+1)|$ and the inequalities 
$E_\text{A}(+1) \geq E_\text{A}(-1)$ and 
$E_\text{B}(+1) \geq E_\text{B}(-1)$
imply that the spectrum $\lambda_{0}$ given by \eqref{eq:lambda0} is reduced) . 
Therefore, it is difficult to reveal the entanglement
and the violation of Bell-CHSH violation with the initial excited
state $|\!\uparrow_{\text{A}}\uparrow_{\text{B}}\rangle$.

\subsection{The initial condition $|\!\downarrow_\text{A} \uparrow_\text{B} \rangle$}

We consider the detectors' initial condition
$|\!\downarrow_\text{A}\uparrow_\text{B} \rangle$. The components $E_\text{A},E_\text{B},E_\text{AB}$ and $X$ of the reduced density matrix are given as 
\begin{align}
E_\text{A}(-1)&=\frac{g_{0}^{2}}{4\pi}  \bigl( e^{-(\Omega \sigma)^{2}}-2\Omega \sigma \,{\rm{Erfc}}[\Omega \sigma] \bigr), \\
E_\text{B}(+1)&=\frac{g_{0}^{2}}{4\pi}  \bigl( e^{-(\Omega \sigma)^{2}}+2\Omega \sigma \,{\rm{Erfc}}[-\Omega \sigma] \bigr), \\
E_\text{AB}(-1,+1)&=\frac{g_{0}^{2}\sigma}{4\pi i r}e^{-2i\Omega t_{0}-(r/2\sigma)^{2}} \Bigl( {\rm{Erfc}} \Bigl[-i\frac{r}{2\sigma} \Bigr] -{\rm{Erfc}} \Bigl[i\frac{r}{2\sigma} \Bigr] \Bigr),  \\
X(-1,+1)&=\frac{g_{0}^{2} \sigma}{4\pi ir}e^{-(r/2\sigma)^{2}} \Bigl( {\rm{Erfc}} \Bigl[-\Omega \sigma-i\frac{r}{2\sigma}\Bigr]+{\rm{Erfc}} \Bigl[\Omega \sigma-i\frac{r}{2\sigma}\Bigr] \Bigr).
\end{align}
The left panel of Fig. \ref{fig9} shows the contour plot of the
negativity with the green dotted line $\beta=1$. Further, we add the
orange line which denotes $\mathcal{N}=0$ for the initial state
$|\!\downarrow_\text{A}\downarrow_\text{B}\rangle$. 
\begin{figure}[htbp]
   \centering
   \includegraphics[width=0.45\linewidth]{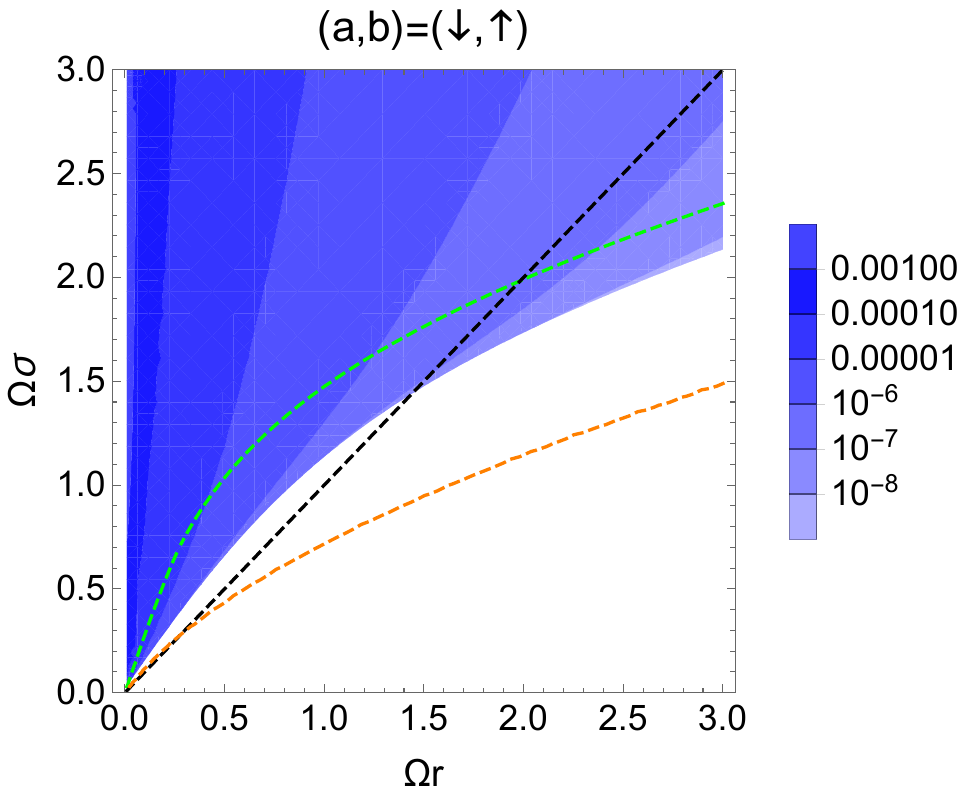}
   \includegraphics[width=0.45\linewidth]{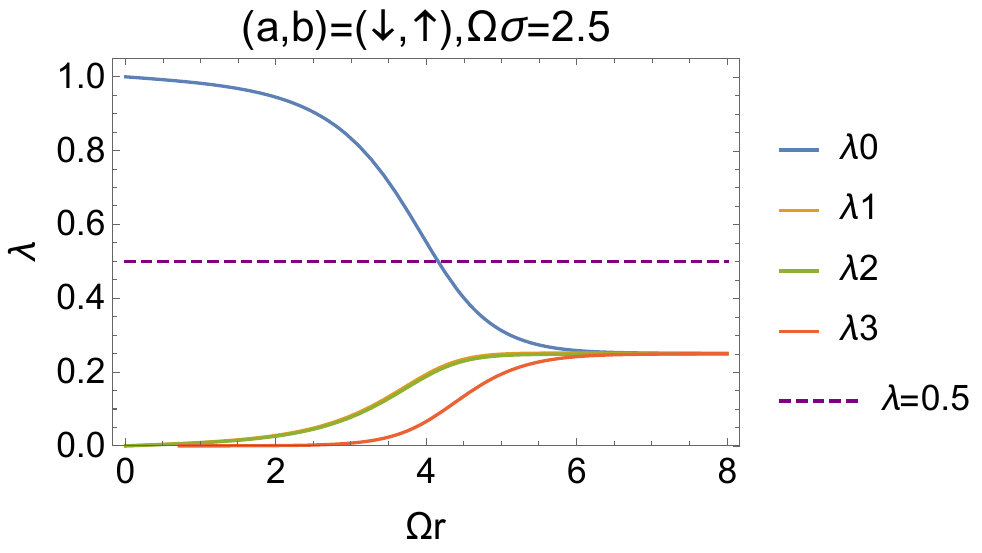}
 \caption{Left panel : The contour plot of the negativity with the
   initial condition
   $|\!\downarrow_\text{A}\uparrow_\text{B}\rangle$. The orange dashed
   line denotes the boundary of the nonzero negativity with the
     initial state
   $|\!\downarrow_\text{A}\downarrow_\text{B}\rangle$. In this case,
   the size of the parameter region which shows the quantum correlations is smaller compared to the case
   $|\!\downarrow_\text{A}\downarrow_\text{B} \rangle$. Right
   panel : The behavior of the spectrum $\{\lambda_{\mu}\}$ as a function of
$\Omega r$ with
   fixed $\Omega \sigma=0.5$ and $g_{0}=10^{-2}$. }
    \label{fig8}
\end{figure}
\noindent
We observe that the region with the nonzero negativity is smaller
compared to that for the initial state
$|\!\downarrow_\text{A}\downarrow_\text{B}\rangle$ but larger than that for the initial state
$|\!\uparrow_\text{A}\uparrow_\text{B}\rangle$. 
This is because the probability of spontaneous emission is large 
($E_\text{B}(+1) \geq E_\text{B}(-1)$) and the generation of quantum correlations  are suppressed compared to the case with 
the initial condition $|\!\downarrow_\text{A}\downarrow_\text{B}\rangle$

\section{Effect of local emissions for the detection region of Bell nolocality and its success probability}

\begin{figure}[htbp]
   \centering
   \includegraphics[width=0.45\linewidth]{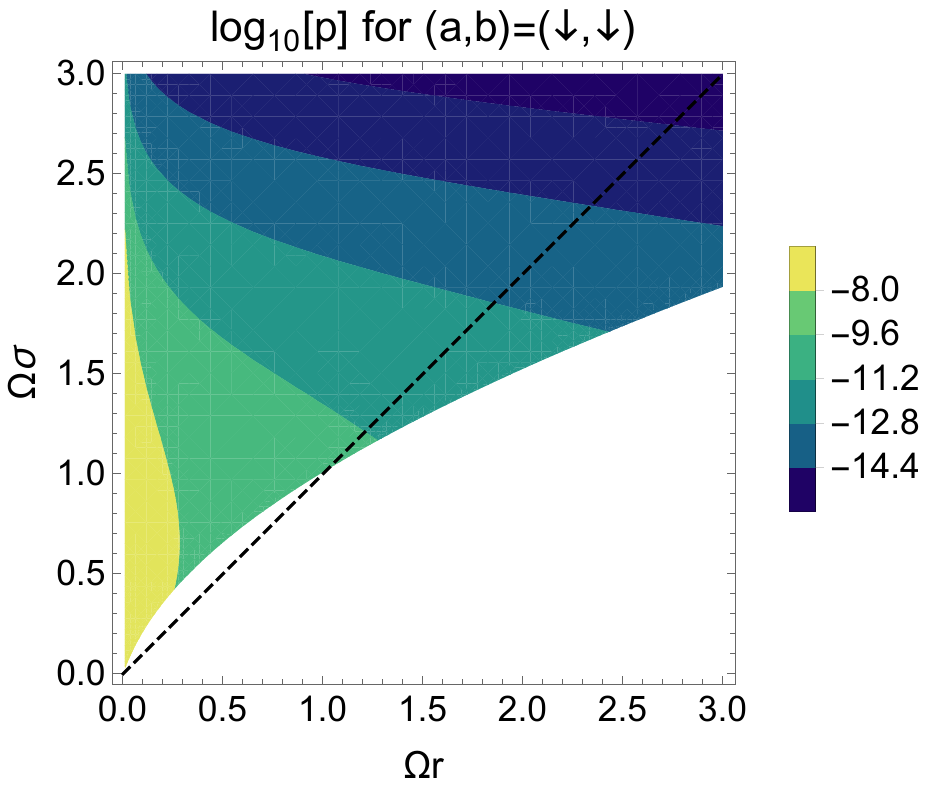}
   \includegraphics[width=0.45\linewidth]{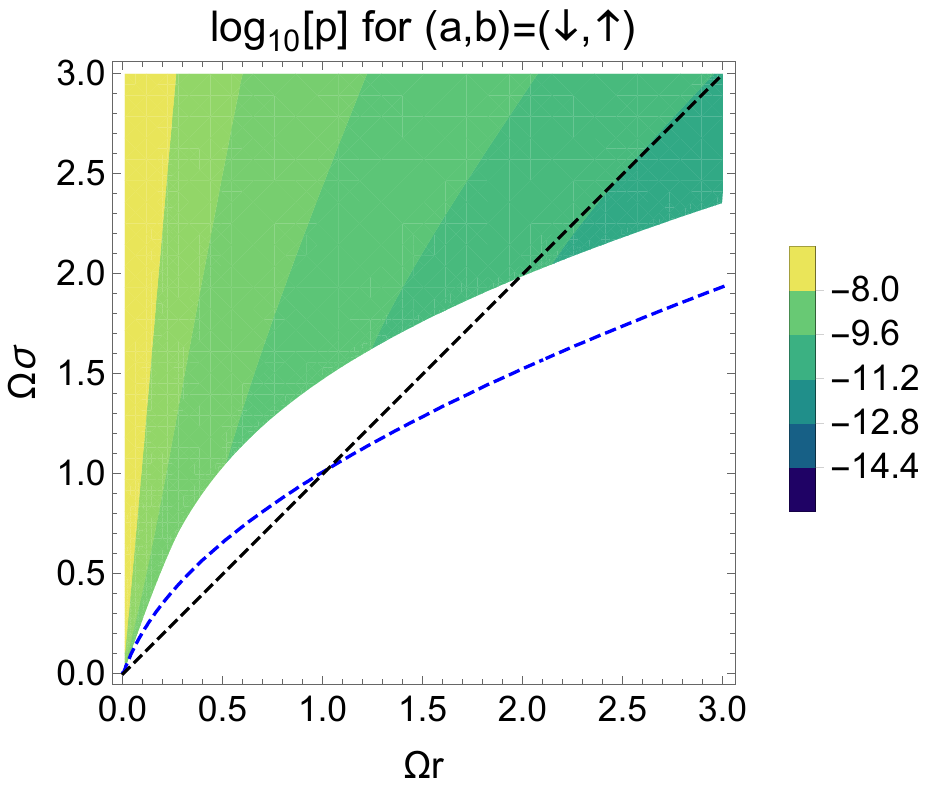}
 \caption{Left panel : The contour plot of the success probability $p$ in common logarithms scale for the initial condition $|\!\downarrow_\text{A} \downarrow_\text{B} \rangle$. The colored region denotes that $\beta$ is larger than unity in the parameter space $(\Omega r, \Omega \sigma)$. As the distance or the interaction time increase, the probability becomes small. Right panel : The contour plot of the success probability $p$ in common logarithms scale with $\beta>1$ for the initial condition $|\!\downarrow_\text{A} \uparrow_\text{B} \rangle$. The blue dotted curve represents the curve $\beta=1$ for the initial condition $|\!\downarrow_\text{A} \downarrow_\text{B} \rangle$ equals to unity. The detection region of the violation of the Bell-CHSH inequality is reduced, but its probability increases than the case for the initial condition $|\!\downarrow_\text{A} \downarrow_\text{B} \rangle$.}
    \label{fig9}
\end{figure}

In this section, we focus on the success probability of the optimal filtering for the initial conditions $|\!\downarrow_\text{A} \downarrow_\text{B} \rangle$ and $|\!\downarrow_\text{A} \uparrow_\text{B} \rangle$. In the left panel of Fig. \ref{fig9}, we present the contour plot of the success probability with the violation of the Bell-CHSH inequality (that is, $\beta>1$) in common logarithms scale for the initial condition 
$|\!\downarrow_\text{A} \downarrow_\text{B} \rangle$ and observe that the probability decreases as the distance between the two detectors or the interaction time increase.
The right panel of Fig. \ref{fig9} is the contour plot of $p$ in common logarithms scale for the initial condition $|\!\downarrow_\text{A} \uparrow_\text{B} \rangle$. Comparing to the case $|\!\downarrow_\text{A} \downarrow_\text{B} \rangle$, the behavior of the probability for the interaction time is different. To understand the behavior of $p$ for $\Omega \sigma \gg 1 $, we should note that the success probability in the leading order of the coupling is 
\begin{equation}
p\approx 2 \sqrt{ E_\text{A}E_\text{B}+|E_\text{AB}|^{2}+|X|^{2}} (\sqrt{E_\text{A}E_\text{B}+|E_\text{AB}|^{2}+|X|^{2}}
+\sqrt{E_\text{A}E_\text{B}}), \label{eq:p-det}
\end{equation}
where we substituted Eqs. (10)-(13) into the formula of the probability \eqref{eq46}. 
Here let us denote $p(-1,-1)$ and $p(-1,+1)$ as the success probability for each initial condition $|\!\downarrow_\text{A} \downarrow_\text{B} \rangle$ and $|\!\downarrow_\text{A} \uparrow_\text{B} \rangle$. For $\Omega \sigma \gg 1$ the switching function $g(t)$ is effectively regarded as the constant $g_{0}$. Hence the total system is invariant under a time translation and the total energy is conserved. By the energy conservation for $\Omega \sigma \gg 1$, each component $E_\text{A}, E_\text{B}, E_\text{AB}$ and $X$ is 
\begin{align}
E_\text{A}(a)&= \int d^{3}k |\langle-a,b,\bm{k}_\phi| \tilde{\Psi}_\text{out} \rangle|^{2} \approx \int d^{3}k M^{2}_\text{A}(a,\bm{k}) \delta(\omega_{\bm{k}}-a\Omega), \\
E_\text{B}(b)&= \int d^{3}k |\langle a,-b,\bm{k}_\phi| \tilde{\Psi}_\text{out} \rangle|^{2} \approx \int d^{3}k M^{2}_\text{B}(b,\bm{k}) \delta(\omega_{\bm{k}}-b\Omega), \\
E_\text{AB}(a,b)&=\int d^{3}k \langle-a,b,\bm{k}_\phi| \tilde{\Psi}_\text{out} \rangle \langle \tilde{\Psi}_\text{out} |a,-b,\bm{k}_\phi \rangle \nonumber \\
&\approx \int d^{3}k M_\text{A}(a,\bm{k})M_\text{B}(b,\bm{k}) \delta(\omega_{\bm{k}}-a\Omega)\delta(\omega_{\bm{k}}-b\Omega), \\
X(a,b) &=\langle -a, -b, 0_\phi|\tilde{\Psi}_\text{out} \rangle \approx M_\text{X}(a,b) \delta(-a\Omega-b\Omega),
\end{align}
where $\omega_{\bm{k}}=|k|$ and $M_\text{A},M_\text{B}, M_\text{X}$ correspond to the amplitudes of each transition process. According to the behaviors of 
$E_\text{A},E_\text{B}, E_\text{AB}$ and 
$X$, we find that the probabilities 
$p(-1, -1)$ and 
$p(-1,+1)$ approach zero and 
$2|X|^{2}$ for a large 
$\Omega \sigma$, respectively. In Fig.\ref{fig9}, this different feature appears as the shape of the contour of 
$p$ for each initial condition. 

In the right panel of Fig. \ref{fig9}, we also observe that the success probability is improved compared to the initial condition 
$|\!\downarrow_\text{A} \downarrow_\text{B} \rangle$ while the detection region of the violation of the Bell-CHSH inequality in the parameter space is reduced. This trade-off relation is explained by the formulas of the eigenvalue 
$\lambda_{0}$ \eqref{eq:lambda0} and the success probability 
$p$ \eqref{eq:p-det}. The eigenvalue 
$\lambda_{0}$ is given by the difference 
$|X|-\sqrt{E_\text{A}E_\text{B}}$, and the success probability depends on the sum 
$|X|^{2}+E_\text{A}E_\text{B}$. As the transition probability 
$E_\text{A}$ or 
$E_\text{B}$ of the spontaneous emission grows, the eigenvalue 
$\lambda_{0}$ decreases and the success probability $p$ increases. This means that the process of the local emission has two effects : one is the reduction of the detection region of the violation of the Bell-CHSH inequality, and the other is the increase of the efficiency for the detection of that violation. The local dynamics of each detector plays an important role for the reliable detection of the Bell-CHSH inequality.


\section{Summary and conclusion}

We investigated the detection of quantum correlations of a massless
scalar field for the model of two qubit detectors. We considered the two qubit detectors coupled to the scalar field in a Minkowski vacuum. Under the second order perturbation of the total dynamics, we examined the negativity and the violation of 
the Bell-CHSH inequality for the detectors' state. It is demonstrated 
that the state of the detectors can be entangled but satisfies the Bell-CHSH inequality
within our perturbative treatment. To reveal the violation of the Bell-CHSH inequality, we used the 
optimal filtering operations acting on each detector. In general it is complicated to construct 
this operations. It is simpler to obtain the optimal filtering for the out-state of the detectors described by an X state. Such a
filtering is given by the two steps; passing through an amplitude damping channel and choosing a specific outcome after the channel. The success probability to perform the filtering characterizes the efficiency of the detection of the violation of Bell-CHSH inequality. 

By examining the negativity and the violation of the Bell-CHSH inequality under the optimal filter, we found that the detection of the quantum correlations strongly depends on the initial state of the detectors. When the detectors are initially in the ground state, the detection region in the parameter space showing the quantum correlation is larger than that obtained for the case of initially excited states. This is because the excited detectors spontaneously emit the scalar particles and such a local dynamics prevents the detection of quantum correlations by the detectors.

Further we focused on the success probability of the optimal filtering
for the violation of Bell-CHSH inequality for the initial condition 
$|\!\downarrow_\text{A} \downarrow_\text{B} \rangle$ and 
$|\!\downarrow_\text{A} \uparrow_\text{B} \rangle$. 
Then we demonstrated the trade-off relation between the parameter region of the violation of Bell-CHSH inequality and the success probability. Due to this trade-off relation, the reliable
detection of the violation of Bell-CHSH inequality becomes non-trivial, and
we found that the setting with the initial state 
$|\!\!\downarrow_\text{A}\uparrow_\text{B}\rangle$ of the detectors leads to the violation of the Bell-CHSH inequality with a
high efficiency and large parameter regions. This result gives the
suitable model for the detection of the violation of Bell-CHSH inequality through the two qubit detectors.

\begin{acknowledgments}
This work was supported in part by the JSPS KAKENHI Grant Number 15K05073.
\end{acknowledgments}

\begin{appendix}
\section{Components of reduced density matrix}
\begin{figure}[htbp]
   \centering
   \includegraphics[width=0.4\linewidth]{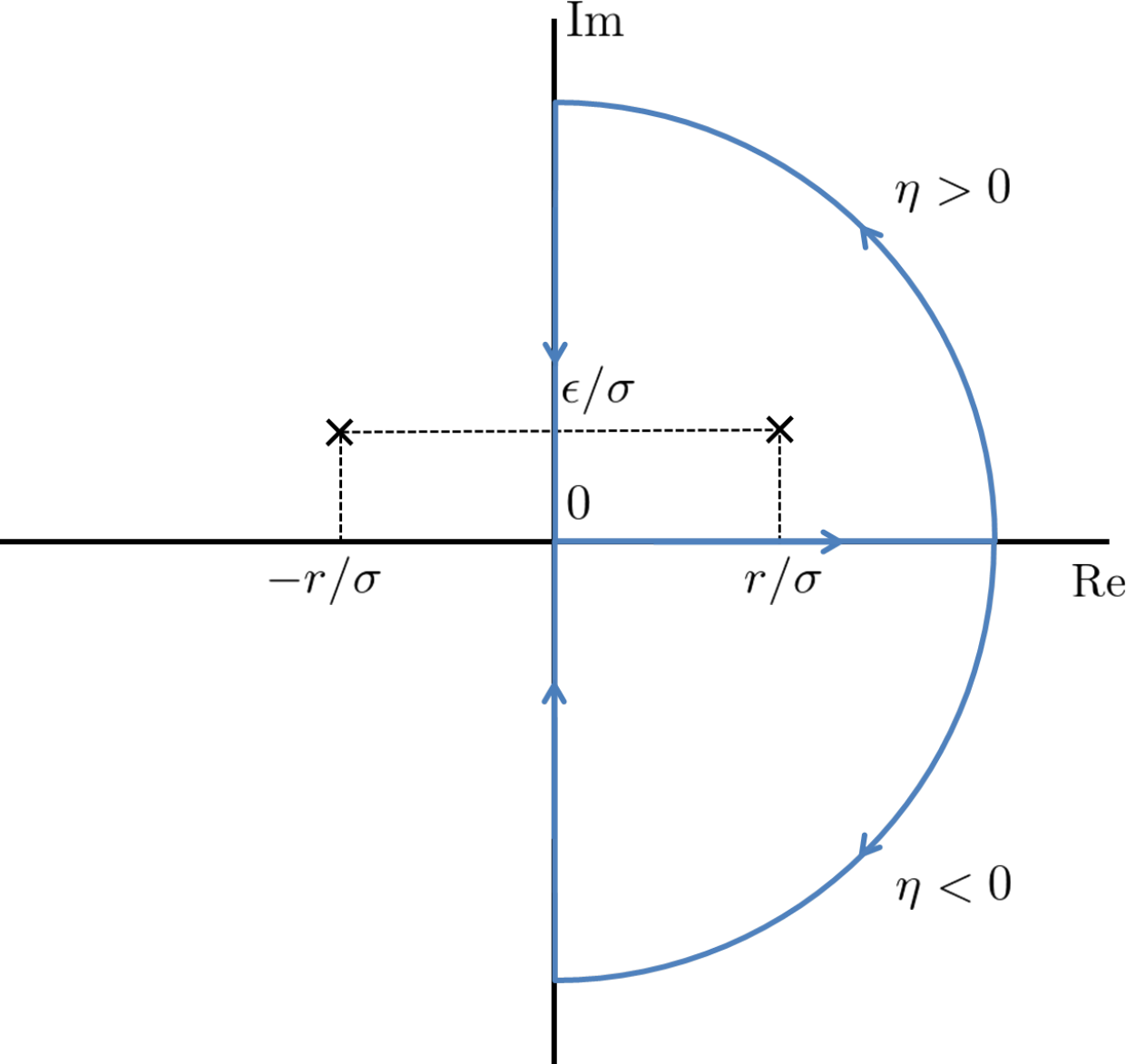}
    \caption{The contour of the complex integration to compute the non-diagonal component $X^{*}$.} 
    \label{fig11}
\end{figure}

The diagonal components $E_\text{A}$ and $E_\text{B}$ are obtained from
$E_\text{AB}$. According to Eq. \eqref{eq:EAB}, the non-diagonal component $E_\text{AB}$ are represented by
\begin{equation}
E_\text{AB}
=\int d^{3}k\,\langle 0_{\phi}| \Phi^\text{A}_{a}|\bm{k}_{\phi} \rangle \langle \bm{k}_{\phi}|\Phi^\text{B}_{-b}|0_{\phi} \rangle,
\end{equation}
where note that the inner product of $\Phi^\text{A}_{a}|0_{\phi} \rangle$
and $n$-particle state for $n \geq2$ or $n=0$ is zero. We introduce
the regularized mode function of the Minkowski vacuum
\begin{equation}
u^{\epsilon}_{\bm{k}}(\bm{x},t)=\frac{ e^{-i\omega_{\bm{k}}(t-i\epsilon/2)+i\bm{k} \cdot \bm{x}}}{(2\pi)^{3/2} \sqrt{2\omega_{\bm{k}}}},
\end{equation}
where $\omega_{\bm{k}}=|\bm{k}|$. The inner product
$\langle 0_{\phi}| \Phi^\text{A}_{a}|\bm{k}_{\phi} \rangle$ are calculated
as
\begin{equation}
\langle 0_{\phi}| \Phi^\text{A}_{a}|\bm{k}_{\phi} \rangle
=\int^{\infty}_{-\infty} dt\, g(t)\,e^{-i\Omega at}\, u^{\epsilon}_{k}(\bm{x}_\text{A},t)
=g_{0}\sqrt{2\pi \sigma^{2}}\, e^{-\frac{\sigma^{2}}{2}(\omega_{\bm{k}}-\Omega a)^{2}-i\Omega at_{0}}\,u_{\bm{k}}^{\epsilon}(\bm{x}_\text{A},t_{0}).
\end{equation}
The component $E_\text{AB}$ is computed as
\begin{align}
E_\text{AB}
&=\int d^{3}k\, \langle 0_{\phi}| \Phi^\text{A}_{a}
  |\bm{k}_{\phi}\rangle \langle
  \bm{k}_{\phi}|\Phi^\text{B}_{-b}|0_{\phi} \rangle \nonumber \\
&=2\pi g_{0}^{2}\sigma^{2}e^{i\Omega(a-b)t_{0}} \int d^{3}k\,
  e^{-\frac{\sigma^{2}}{2}(\omega_{\bm{k}}-\Omega
  a)^{2}-\frac{\sigma^{2}}{2}(\omega_{\bm{k}}-\Omega b)^{2}}
  u_{\bm{k}}^{\epsilon}(\bm{x}_\text{A},t_{0})\,
  u_{\bm{k}}^{\epsilon*}(\bm{x}_\text{B},t_{0}) \nonumber \\
&=\frac{g_{0}^{2}\sigma}{4i\pi r}e^{i\Omega(a-b)t_{0}}
  \int^{\infty}_{0} du\, e^{-\frac{1}{2}(u-\Omega \sigma
  a)^{2}-\frac{1}{2}(u-\Omega \sigma b)^{2}}
  (e^{iur/\sigma}-e^{-iur/\sigma})e^{-\epsilon u/\sigma } \nonumber \\
&=\frac{g_{0}^{2}\sigma}{4i\pi
  r}e^{i\Omega(a-b)t_{0}-(\Omega\sigma)^{2}} \Bigl(\exp
  \Bigl[\bigl(-\frac{\Omega
  \sigma}{2}(a+b)-i\frac{r}{2\sigma}+\frac{\epsilon}{2\sigma}\bigr)^{2}
  \Bigr] {\rm{Erfc}} \Bigl[-\frac{\Omega
  \sigma}{2}(a+b)-i\frac{r}{2\sigma}+\frac{\epsilon}{2\sigma} \Bigr]
  \nonumber \\
&-\exp \Bigl[\bigl(-\frac{\Omega \sigma}{2}(a+b)+i\frac{r}{2\sigma}+\frac{\epsilon}{2\sigma}\bigr)^{2} \Bigr] {\rm{Erfc}} \Bigl[-\frac{\Omega \sigma}{2}(a+b)+i\frac{r}{2\sigma}+\frac{\epsilon}{2\sigma} \Bigr] \Bigr).
\end{align}
We get Eq. \eqref{eq16} by taking the limit
$\epsilon \rightarrow 0$. Next we derive the formula of
$X$. Using the Wightman function
$D^{+}(x-x',t-t')=\langle 0_{\phi}|\phi(x,t)\phi(x',t')| 0_{\phi}
\rangle$
given by Eq. \eqref{eq13}, the non-diagonal component
$X^{*}$ is
\begin{align}
X^{*}
&=-\langle 0^{\phi}| {\rm{T}}[\Phi^\text{A}_{-a}\Phi^\text{B}_{-b}]
  |0^{\phi}\rangle \nonumber \\
&=-\int^{\infty}_{-\infty} dt_{2}\int^{\infty}_{-\infty} dt_{1}\, g(t_{2})g(t_{1})\,e^{i\Omega(-at_{2}-bt_{1})} (\theta(t_{2}-t_{1})D^{+}(r,t_{2}-t_{1})+\theta(t_{1}-t_{2})D^{+}(r,t_{1}-t_{2})) \nonumber \\
&=-2\sqrt{\pi}g_{0}^{2}\sigma^{2}(\Omega \sigma) e^{-i\Omega (a+b)t_{0}-(\Omega \sigma)^{2}} \nonumber \\
&\times \int^{\infty}_{-\infty}dy\, e^{-(\Omega \sigma)^{2}(y+i (a-b)/2)^{2}}(\theta(-y)D^{+}(r,-2\Omega \sigma^{2}y)+\theta(y)D^{+}(r,2\Omega \sigma^{2}y)), \label{A5}
\end{align}
where the integral variables $t_{1}$ and $t_{2}$ are changed as 
\begin{equation}
\Omega \sigma^{2}x=\frac{(t_{1}-t_{0})+(t_{2}-t_{0})}{2},\quad
\Omega \sigma^{2}y=\frac{(t_{1}-t_{0})-(t_{2}-t_{0})}{2}
\end{equation}
and we carried out the $x$ integration. By using the identity
\begin{equation}
e^{-y^{2}}=\frac{1}{\sqrt{\pi}} \int^{\infty}_{-\infty} d\eta~e^{-\eta^{2}+2i\eta y},
\end{equation}
the above equation \eqref{A5} can be rewritten as  
\begin{align}
X^{*}
&=-2g_{0}^{2}\sigma^{2}(\Omega \sigma) e^{-i\Omega (a+b)t_{0}-(\Omega \sigma)^{2}}  \nonumber \\
&\times
\int^{\infty}_{-\infty}d\eta\, e^{-\eta^{2}} \Bigl(e^{-(a-b)(\Omega
  \sigma)\eta }+e^{(a-b)(\Omega \sigma)\eta }
  \Bigr)\int^{\infty}_{0}dy\, e^{2i(\Omega \sigma)\eta y
  }D^{+}(r,2\Omega \sigma^{2}y) \nonumber \\
&=\frac{g_{0}^{2}}{4\pi^{2}}e^{-i\Omega (a+b)t_{0}-(\Omega \sigma)^{2}}  
\int^{\infty}_{-\infty}d\eta\, e^{-\eta^{2}} \Bigl(e^{-(a-b)(\Omega \sigma)\eta }+e^{(a-b)(\Omega \sigma)\eta } \Bigr)\int^{\infty}_{0}dy\,\frac{e^{i\eta y}}{(y-i\epsilon/\sigma)^{2}-(r/\sigma)^{2}} .
\end{align}
The $y$ integration is equivalent to the complex integration given in Fig. \ref{fig11}. Hence, 
\begin{align}
\int^{\infty}_{0}dy \frac{e^{i\eta y}}{(y-i\epsilon/\sigma)^{2}-(r/\sigma)^{2}}
&=\Bigl[\frac{i\pi \sigma}{r} e^{i\eta(r/\sigma+i\epsilon/\sigma)}-i\int^{\infty}_{0}\frac{e^{-\eta y}}{(y-\epsilon/\sigma)^{2}+(r/\sigma)^{2}} \Bigr]\theta(\eta) \nonumber \\
&+i\int^{0}_{-\infty}\frac{e^{-\eta y}}{(y-\epsilon/\sigma)^{2}+(r/\sigma)^{2}} \Bigr]\theta(-\eta), \label{A9}
\end{align}
where the second and third terms are the integration along the
imaginary axis. For $\epsilon \rightarrow 0$ the sum of those terms is
an odd function, and then it does not contribute to the $\eta$
integration (note that the function of $\eta$ in front of Eq.
\eqref{A9} is an even function).
Thus we get the following formula
\begin{align}
X^{*}
&=\frac{g_{0}^{2}}{4\pi^{2}}e^{-i\Omega (a+b)t_{0}-(\Omega \sigma)^{2}}   
\int^{\infty}_{0}d\eta\,  e^{-\eta^{2}} \Bigl(e^{-(a-b)(\Omega \sigma)\eta }+e^{(a-b)(\Omega \sigma)\eta } \Bigr)\frac{i\pi \sigma}{r} e^{i\eta r/\sigma} \nonumber \\
&=\frac{ig_{0}^{2} \sigma}{4\pi r}e^{-i\Omega (a+b)t_{0}-(\Omega \sigma)^{2}}   
\Bigl(\exp \Bigl[\Bigl(\frac{\Omega \sigma}{2}(a-b)+i\frac{r}{2\sigma}\Bigr)^{2}\Bigr] {\rm{Erfc}} \Bigl[-\frac{\Omega \sigma}{2}(a-b)-i\frac{r}{2\sigma}\Bigr] \nonumber \\
&+\exp \Bigl[\Bigl(-\frac{\Omega \sigma}{2}(a-b)+i\frac{r}{2\sigma}\Bigr)^{2}\Bigr] {\rm{Erfc}} \Bigl[\frac{\Omega \sigma}{2}(a-b)-i\frac{r}{2\sigma} \Bigr] \Bigr).
\end{align}

\section{Equality of $|X(-1,-1)|$ and $|X(+1,+1)|$}
Let us show the equality of $|X(-1,-1)|$ and
$|X(+1,+1)|$. Under the second order of the coupling, the
non-diagonal components $X(-1, -1)$ and $X(+1,+1)$ is
\begin{align}
X(-1,-1)
&=\langle \uparrow_\text{A}\uparrow_\text{B}, 0_\phi| \tilde{\Psi}_\text{out} \rangle^{*} \nonumber \\
&= -\int^{\infty}_{-\infty}dt_{1}\int^{t_{1}}_{-\infty}dt_{2}\langle \downarrow_\text{A}\downarrow_\text{B}| \langle 0_{\phi}|\tilde{V}(t_{2})\tilde{V}(t_{1})
  |0_{\phi} \rangle |\!\uparrow_\text{A}\uparrow_\text{B} \rangle \nonumber \\
&= -2  \int^{\infty}_{-\infty}dt_{1}\int^{t_{1}}_{-\infty}dt_2\, g(t_{1}+t_{0})g(t_{2}+t_{0}) e^{-i\Omega(t_{1}+t_{2})} \langle 0_{\phi}|\phi(\bm{x}_\text{B}-\bm{x}_\text{A},t_{2}-t_{1})\phi(0) |0_{\phi} \rangle,  \label{B1}\\
X(+1,+1)
&=\langle \downarrow_\text{A}\downarrow_\text{B}, 0_\phi| \tilde{\Psi}_\text{out} \rangle^{*} \nonumber \\
&= -\int^{\infty}_{-\infty}dt_{1}\int^{t_{1}}_{-\infty}dt_{2}\langle \uparrow_\text{A}\uparrow_\text{B}| \langle 0_{\phi}|
  \tilde{V}(t_{2})\tilde{V}(t_{1}) |0_{\phi} \rangle
  |\downarrow_\text{A}\downarrow_\text{B} \rangle \nonumber \\
&= -2  \int^{\infty}_{-\infty}dt_{1}\int^{t_{1}}_{-\infty} dt_{2}\,e^{i\Omega(t_{1}+t_{2})} g(t_{1})g(t_{2})\langle 0_{\phi}|\phi(\bm{x}_\text{B}-\bm{x}_\text{A},t_{2}-t_{1})\phi(0) |0_{\phi} \rangle ,
\end{align}
where we used the translation invariant of the vacuum state for Eq. \eqref{B1}. Due to the
time reversal invariance of the Minkowski vacuum, $X(-1,-1)$
is rewritten as
\begin{equation}
X(-1,-1)= -2\int^{\infty}_{-\infty}dt_{1}\int^{\infty}_{t_{1}} dt_{2}\, g(-t_{1}+t_{0})g(-t_{2}+t_{0})\, e^{i\Omega(t_{1}+t_{2})-2i\Omega t_{0}} \langle 0_{\phi}|\phi(\bm{x}_\text{B}-\bm{x}_\text{A},t_{2}-t_{1})\phi(0) |0_{\phi} \rangle.
\end{equation}
The switching function $g(t)$ is a Gaussian function, and
$g(t+t_{0})=g(-t+t_{0})$ holds. Thus,
\begin{align}
X(-1,-1)&= -2\int^{\infty}_{-\infty}dt_{1}\int^{\infty}_{t_{1}} dt_{2}\, g(t_{1}+t_{0})g(t_{2}+t_{0})\, e^{i\Omega(t_{1}+t_{2})-2i\Omega t_{0}} \langle 0_{\phi}| \phi(\bm{x}_\text{B}-\bm{x}_\text{A},t_{2}-t_{1})\phi(0) |0_{\phi} \rangle \nonumber \\
&= -2\int^{\infty}_{-\infty}dt_{1}\int^{\infty}_{t_{1}}dt_{2}\, g(t_{1})g(t_{2})\, e^{i\Omega(t_{1}+t_{2})-4i\Omega t_{0}} \langle 0_{\phi}|\phi(\bm{x}_\text{B}-\bm{x}_\text{A},t_{2}-t_{1})\phi(0) |0_{\phi} \rangle \nonumber \\
&=e^{-4i\Omega t_{0}}X(+1,+1),
\end{align}
and we get the equality
$|X(-1,-1)|=|X(+1,+1)|$.
\end{appendix}


\end{document}